\UseRawInputEncoding
\documentclass[times,3p,review,preprint,10pt]{elsarticle}

\usepackage{citesort}
\usepackage{url}
\usepackage[fleqn]{amsmath}  
\usepackage{amssymb}
\usepackage[colorlinks,
linkcolor=cyan,
urlcolor=cyan,
citecolor=cyan]{hyperref}
\usepackage{graphicx} 
\usepackage{float} 
\usepackage{subfigure} 
\usepackage{amsfonts}
\usepackage{stackrel}
\usepackage[boxed,ruled,lined,commentsnumbered]{algorithm2e}
\usepackage{algpseudocode}
\usepackage{booktabs}
\usepackage{multirow}
\usepackage{caption}
\usepackage{epstopdf}
\usepackage{xcolor}
\usepackage {ragged2e}
\begin{document}
\begin{frontmatter}



\title{Multi-modal Evidential Fusion Network for Trustworthy PET/CT Tumor Segmentation} 


\author[1]{Yuxuan Qi\fnref{label1}}
\ead{20225228030@stu.suda.edu.cn}

\author[2,3]{Li Lin\fnref{label1}\corref{cor1}}
\ead{linli@eee.hku.hk}
\fntext[label1]{Yuxuan Qi and Li Lin contributed equally to this study.}
\address[1]{School of Electronic Information, Soochow University,Suzhou, 215031, China}

\address[2]{Department of Electrical and Electronic Engineering, the University of Hong Kong, Hong Kong,
999077,China}

\address[3]{Department of Electronic and Electrical Engineering, Southern University of Science and Technology, Shenzhen, 518055, China}
\author[4]{Jingya Zhang}
\ead{zhangjy0611@163.com}
\address[4]{School of Electronic and Information Engineering, Changshu Institute of Technology, Suzhou, 215506, China}
\author[5]{Bin Zhang}
\address[5]{Department of Nuclear Medicine, the First Affiliated Hospital of Soochow University, Suzhou 215006, Peoples Republic of China}
\ead{zbnuclmd@126.com}

\author[1]{Jiajun Wang\corref{cor1}}
\ead{jjwang@suda.edu.cn}
\cortext[cor1]{Corresponding author.}


\begin{abstract}
  Accurate tumor segmentation in PET/CT images is crucial for computer-aided cancer diagnosis and treatment. The primary challenge lies in effectively integrating the complementary information from PET and CT images. In clinical settings, the quality of PET and CT images often varies significantly, leading to uncertainty in the modality information extracted by networks. To address this challenge, we propose a novel Multi-modal Evidential Fusion Network (MEFN), which consists of two core stages: Cross-Modal Feature Learning (CFL) and Multi-modal Trustworthy Fusion (MTF). The CFL stage aligns features across different modalities and learns more robust feature representations, thereby alleviating the negative effects of domain gap. The MTF stage utilizes mutual attention mechanisms and an uncertainty calibrator to fuse modality features based on modality uncertainty and then fuse the segmentation results under the guidance of Dempster-Shafer Theory. Besides, a new uncertainty perceptual loss is introduced to force the model focusing on uncertain features and hence improve its ability to extract trusted modality information. Extensive comparative experiments are conducted on two publicly available  PET/CT  datasets to evaluate  the performance of our proposed method whose results  demonstrate that our MEFN significantly outperforms state-of-the-art methods with improvements of   3.10$\%$ and 3.23$\%$ in DSC scores on the AutoPET dataset and the Hecktor dataset, respectively. More importantly, our model can provide radiologists with credible uncertainty  of the segmentation results  for their decision in accepting or rejecting the automatic segmentation results, which is particularly important for clinical applications. Our code will be available at \href{https://github.com/QPaws/MEFN}{https://github.com/QPaws/MEFN}.
  \end{abstract}




\end{frontmatter}


\section{Introduction}
  PET (Positron Emission Tomography) and CT (Computed Tomography) serve as two major medical imaging techniques that play  crucial roles in cancer diagnosis and treatment. CT images provide high-resolution anatomical information for tumor edge detection and hence many researchers in recent years \citep{li2018,yu2019} developed algorithms  for the segmentation of tumor regions in the liver from  CT images. However, these methods often face challenges in accurately locating and delineating tumors due to tissue density variations, noise, and limited contrast, especially for small or subtle lesions. PET images, on the other hand, measure the level of metabolic activity which reflects the functional status of tissues and cells and offers advantages for disease diagnosis and monitoring. Nevertheless, their lower resolution can lead to relatively blurred tumor boundaries and anatomical structures, which makes it difficult to meet the demands for segmentation accuracy. The complementarity of the CT and PET pushes the simultaneous examination with these two imaging modalities, which urges the development of joint tumor segmentation algorithms from PET/CT multi-modal images \citep{ZHU2024}.

 In radiology, tumor segmentation plays a critical role in accurate diagnosis and treatment.  The most critical challenge in multimodal segmentation is how to fuse the information from multiple modalities. Existing multimodal fusion methods can be  categorized into three types: input-level fusion, feature-level fusion, and decision-level fusion. In input-level fusion, PET and CT images are typically fused by concatenation \citep{zhang2015,chen2018}. However, PET/CT joint segmentation requires capturing complementary information between the two modalities but this fusion approach often struggles to explore nonlinear relationships between modalities, mak-eing it difficult to fully utilize the inter-modality information. In feature-level fusion, high-dimensional features from different modalities are typically extracted separately and then combined together. However, simple feature fusion \citep{LI2023} will introduce redundant or conflicting features, which can degrade network performance. Besides, the existence of a domain gap and the significant feature differences between PET and CT modalities, unlike the relatively consistent features in MR/CT, further complicate the process of effectively fusing information from these two modalities within a network. In decision-level fusion, researchers often use different segmentation networks to segment tumors in individual modalities separately \citep{nie2016}. The segmentation results of each modality are then fused using specific fusion strategies to obtain the final segmentation result. However, in practical clinical applications, the quality of multi-modal data is often unstable (e.g., some modalities may be corrupted), so the quality of multi-modal inputs should be quantitatively measured in some way to provide doctors with predicting confidence, which is particularly important when deploying multi-modal segmentation models for clinical tasks \citep{GUARRASI2024}.

This paper aims to address the aforementioned issues and proposes a Multi-modal Evidential Fusion Network (MEFN) to achieve more accurate and reliable PET/CT multi-modal tumor segmentation. Specifically, this network comprises two training stage: the Cross-Modal Feature Learning (CFL) stage based on modality translation and multi-task learning for more robust representation learning and the Multi-modal Trustworthy Fusion (MTF) stage for multimodal tumor segmentation. These two stages are alternately trained to improve feature representation and segmentation accuracy. In CFL stage, we use Generative Adversarial Networks (GAN) to enable transformation between the two modalities and set additional tumor decoders to guide the network to focus on tumor features. Then, the trained network parameters in CFL are shared with the segmentation backbone of MTF and as the initial weights of the network during training. Such modality translation aligns features across different modalities into a shared feature space, enhancing the model's ability to capture relevant features and improving segmentation performance with a more robust multimodal representation during the MTF stage. In MTF stage, in addition to the segmentation backbone, we also introduce the Dual-attention Feature Calibrating (DFC) module and adopt the Dempster-Shafer Theory-based Trustworthy Fusion (DTF) method. These two methods respectively address feature-level fusion and decision-level fusion, considered the uncertainty between different modalities and reduced the impact of redundant and conflicting features on the model.

In summary, this work makes the following main contributions:
\begin{itemize}
    \item A GAN-based modality translation and multi-task constrained CFL stage is proposed to obtain more robust representation of the model to alleviate the impact of domain gap between different modalities.
    \item A feature-level fusion method named DFC is introduced to provide a more reasonable weight allocation for feature fusion by reducing the impact of redundant and conflicting features.
    \item To handle the quality differences between the modalities, a decision-level fusion strategies is proposed to allocate fusion weights based on the uncertainty between different modalities in an interpretable manner.
    \item Extensive experiments on two large PET/CT public datasets demonstrate that the proposed method outperforms state-of-the-art multi-modal segmentation algorithms. Ablative experiments are conducted to demonstrate the effectiveness and robustness of the individual modules proposed in our work from different perspectives.
\end{itemize}
\label{sec:Introduction}
\section{Related work}
\subsection{Generative adversarial model}
Generative Adversarial Networks (GAN) are based on zero-sum games in game theory and consist of a generator and a discriminator \citep{goodfellow2020}. The generator creates data by learning its original distribution while the discriminator differentiates between real and generated data and hence guides the generator toward more realistic outputs. GAN is often used for data reconstruction \citep{wolterink2017,wang2018}, synthetic data generation \citep{bowles2018} and converting data between modalities \citep{ben2019}. For example, upon incorporating tumor segmentation labels into the generator, Bi et al. \citep{bi2017} attempted to synthesize PET from CT using a conditional GAN (cGAN). However, this method is impractical for us as segmentation labels are the target predictions in our task. Zhang et al. \citep{xiang2022} proposed Modality-Specific Segmentation Networks with integrated cGAN to extract common features from PET and CT images for lung tumor segmentation. However, this method merges image synthesis with segmentation tasks, which makes the computational overhead high. Ding et al. \citep{DING2024} proposed a cross-modality transfer learning to allow knowledge learned from one imaging modality to be transferred to improve segmentation performance on another modality. This method, however, only considers one-way conversion. Motivated by the aforementioned method, we propose the Cross-Modal Feature Learning (CFL) stage where the Tumor Guided Attention (TGA) mechanism is introduced to help the model focus on tumor regions while simultaneously learning more robust representations and alleviating the impact of the domain gap.
\label{sec:Generative adversarial model}
\subsection{Uncertainty estimation}
In clinical practice, AI often lacks sufficient decision-making information, which makes it crucial to quantify uncertainty for safe application \citep{begoli2019}. Current research on uncertainty estimation includes Bayesian \citep{gal2016}, ensemble \citep{lakshminarayanan2017}, deterministic \citep{Amersfoort2020} and evidence-based methods \citep{huang2022}. Bayesian methods in neural networks replace fixed parameters with probability distributions, leading to probabilistic output labels. For example, Monte Carlo dropout (MC dropout) is introduced as a Bayesian approximation \citep{gal2016}, where dropout layers are used in multiple inference runs to mimic a Gaussian process. Although this strategy is computationally less demanding, it leads to inconsistent outputs \citep{kohl2018}. Ensemble methods improve neural network classification by using model averaging for uncertainty estimation, involving training multiple models with random initialization and adversarial training \citep{kamnitsas2018,lakshminarayanan2017}. While this method doesn't require altering the network architecture, it necessitates retraining models from scratch and hence incurs high computational costs for complex models. Deterministic-based methods, like Deep deterministic uncertainty \citep{mukhoti2021}, have been extended to semantic segmentation using feature space density. However, they alter network structure and increase computational costs. A recent study \citep{huang2022} suggested using deep feature extraction and evidence layers for segmenting lymphoma in PET and CT scans. Rather than  to obtain more robust segmentation by calibrating the uncertainty, these studies tried to to improve the performance of segmentation upon guiding the the uncertainty. They only utilized the generated uncertainty to evaluate the segmentation results but not to  further optimize the network training. Instead, our approach introduces not only an Uncertainty Perceptual Loss for robust segmentation by forcing the network focusing on the prediction results of regions with large uncertainty during training  but also an Uncertainty Calibrator (UC) to further refine network training by calibrating the extracted semantic information according to uncertainty.
\label{sec:Uncertainty quantification}
\subsection{Multi-modal fusion}
Multi-modal medical image segmentation, essential for accurately identifying tumors, employs various fusion strategies: input-level, feature-level and decision-level fusion. A typical example of input-level fusion is the 3D-Inception ResNet \citep{qayyum2021}, which combined PET and CT images in channel-wise manner. Some recent studies \citep{wang2021,liu2021}  employed attention mechanisms in their models to allocate varying weights to different modality spaces and  achieved fairly better results. Although the input-level fusion preserves the original details of the image, it fails in utilizing of the interrelations between different modalities. Feature-level fusion trains networks with individual modal images and then fuses  features of different modalities across layers for final segmentation. It can effectively integrate and utilize multi-modal images \citep{ZHOU2023}. Most existing methods in literature fuse multi-modal features by directly combining features of each modality \citep{WANG2023Cascaded}, which potentially miss key information and introduce redundancies and hence hinder network performance. Ahmad et al. \citep{ahmad2023} introduced the Anatomy Aware Tumor Segmentation Network (AATSN), which uses an encoder to extract CT anatomical and an encoder to extract PET metabolic features for a lightweight fusion attention decoder. In this way, the decoder can use CT for spatial and PET for metabolic reference and hence improves tumor segmentation. However, its lightweight attention mechanism might under-perform with complex data due to limited training parameters. Decision-level fusion employs separate networks for each modality image. Each network uses its modality specific information whose outputs are integrated for the final result. Conventional fusion strategies include averaging and majority voting. Kamnitsas et al. \citep{kamnitsas2018} used an averaging strategy where the authors trained three networks and averaged their confidence scores for voxel segmentation. The majority voting strategy assigns voxels the most frequent label from all networks. However, these methods often overlook the uncertainty in each modality and lead to potential errors in probabilistic fusion, which is fatal in clinical practice. Huang et al. \citep{huang2021} combined PET and CT results using a Dempster-Shafer Theory-based fusion layer in 3D UNet. But this method, limited to decision layer fusion, shows less efficiency in handling multi-modal information. To resolve this problem, this paper merges feature-level and decision-level fusion strategies with the Dual-attention Feature Calibrating (DFC) module and Dempster-Shafer Theory-based Trustworthy Fusion (DTF) method, respectively. DFC reduces feature redundancies  and calibrates low-confidence outputs. It's worth noting that, differing from Huang et al. \citep{huang2021}, our DTF fuses PET and CT data based on uncertainty and mutual information, which greatly improves multi-modal fusion efficiency.
\label{sec:Multimodal fusion}
\label{sec:Related work}

\begin{figure*}[t]
    \centering
    \includegraphics[width=\textwidth]{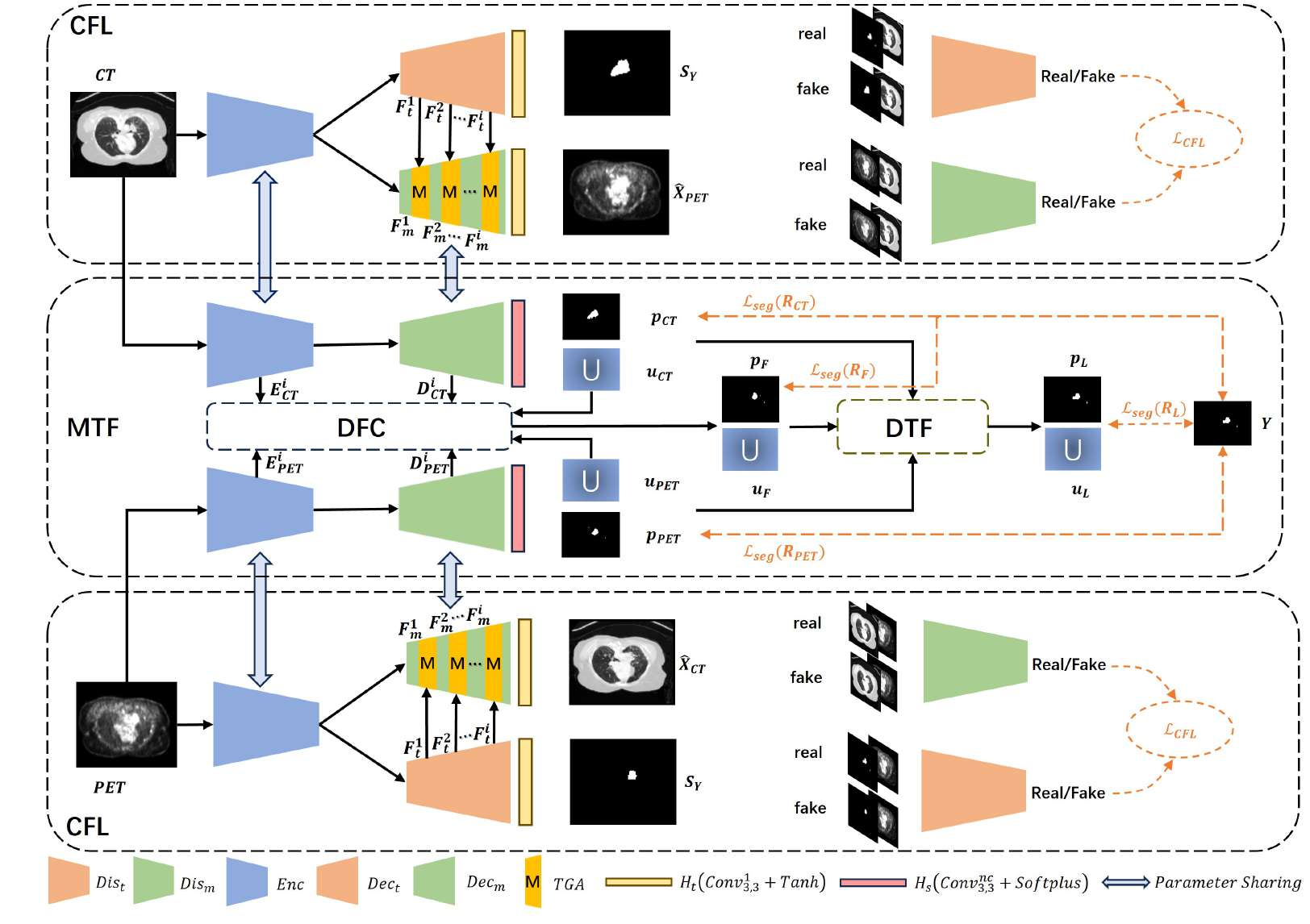}
    \caption{\justifying Framework of our proposed MEFN which consists of CFL stage and MTF stage, both with two branches for PET and CT, respectively. For simplicity, $Enc$ represents the encoder $Enc^{CT}$ or $Enc^{PET}$, $Dec_{m}$ represents the modality decoder $Dec_{m}^{CT}$ or $Dec_{m}^{PET}$, $Dec_{t}$ represents the tumor decoder $Dec_{t}^{CT}$ or $Dec_{t}^{PET}$, $Dis_{m}$ represents the modality discriminator $Dis_{m}^{CT}$ or $Dis_{m}^{PET}$, $Dis_{t}$ represents the shared tumor discriminator.}
    \label{fig:Overview}
\end{figure*}

\section{Methodology}
In this section, we commence by succinctly outlining the MEFN framework, before delving into a comprehensive exposition of its three pivotal components: the Cross-Modal Feature Learning (CFL) stage, the Dual-attention Feature Calibrating (DFC) module and the Dempster-Shafer Theory Based Trustworthy Fusion (DTF). The overall framework of MEFN is provided in Fig. \ref{fig:Overview}.

\subsection{Overview of MEFN}
Defining CT image as $X_{CT}$ and PET image as $X_{PET}$, the two inputs ($X_{CT}$ and $X_{PET}$)  are first fed into the CFL. Within the CFL, it orchestrates the generation of pseudo PET images $\hat{X}_{PET}$ and pseudo CT images $\hat{X}_{CT}$ for each modality through adversarial training. Concurrently, an auxiliary tumor decoder branch $Dec_{t}$ produces corresponding artificial tumor mask images $S_{Y}$. In the training phase, the network acquires modality-specific features of CT and PET along with common tumor features, thereby learning more robust feature representations. Subsequently, the network parameters trained in the CFL are shared to the segmentation backbone as the initial weights of the network. Next, the encoder and decoder features from the two segmentation backbones are fed into the DFC. Within the DFC, encoder and decoder features undergo two distinct fusion strategies: in terms of Cross Modality Attention (CMA) and in terms of Uncertainty Calibrator (UC), respectively. The former adaptively adjusts features extracted by the encoder through mutual attention, which significantly reduces feature redundancies and enhances fusion efficiency. The latter calibrates semantic features extracted by the network using uncertainty factors, which forces the network focusing on areas of high uncertainty within modalities and thus diminishes feature confliction between modalities and enhancing fusion accuracy. Features refined by CMA and UC are then input into the fusion decoder $Dec_{f}$ to obtain the fused segmentation results $Y_{F}$ and $U_{F}$ after upsampling. Finally, segmentation results from the CT branch ($Y_{C}$ and $U_{C}$), PET branch ($Y_{P}$ and $U_{P}$), along with the fused segmentation results ($Y_{F}$ and $U_{F}$), are collectively processed in terms of DTF for decision-level fusion. In DTF, the segmentation results of each branch will dynamically adjust the fusion weight according to the uncertainty, making the fusion result more robust, then the final segmentation outcomes can be acquired as $\hat{Y}$ and $U_{L}$. Although the network structures in CFL and MTF are not necessarily the same, the more network layers these two networks share, the richer the modality-specific features can be transferred from CFL to MTF. To this end, we have designed two stage with quite similar network structures and the only difference is the number of kernels set to the decoder's output head. As shown in Fig. \ref{fig:Overview}, the decoder's output head in CFL ($H_t$) uses one convolutional kernels and employs a $Tanh$ activation function, while the number of convolutional kernels in  the decoder's output head in MTF ($H_s$) corresponds to the number of segmentation classes and  a $Softplus$ activation function is utilized in this case. As can be seen, such an MTF network architecture  could share the most network layers with the CFL and thus can take full advantage of the features learned from the CFL. 
\label{sec:MEFN}

\subsection{Cross-modality Feature Learning (CFL)}
\begin{figure*}[htbp]
    \centering
    \includegraphics[width=\textwidth]{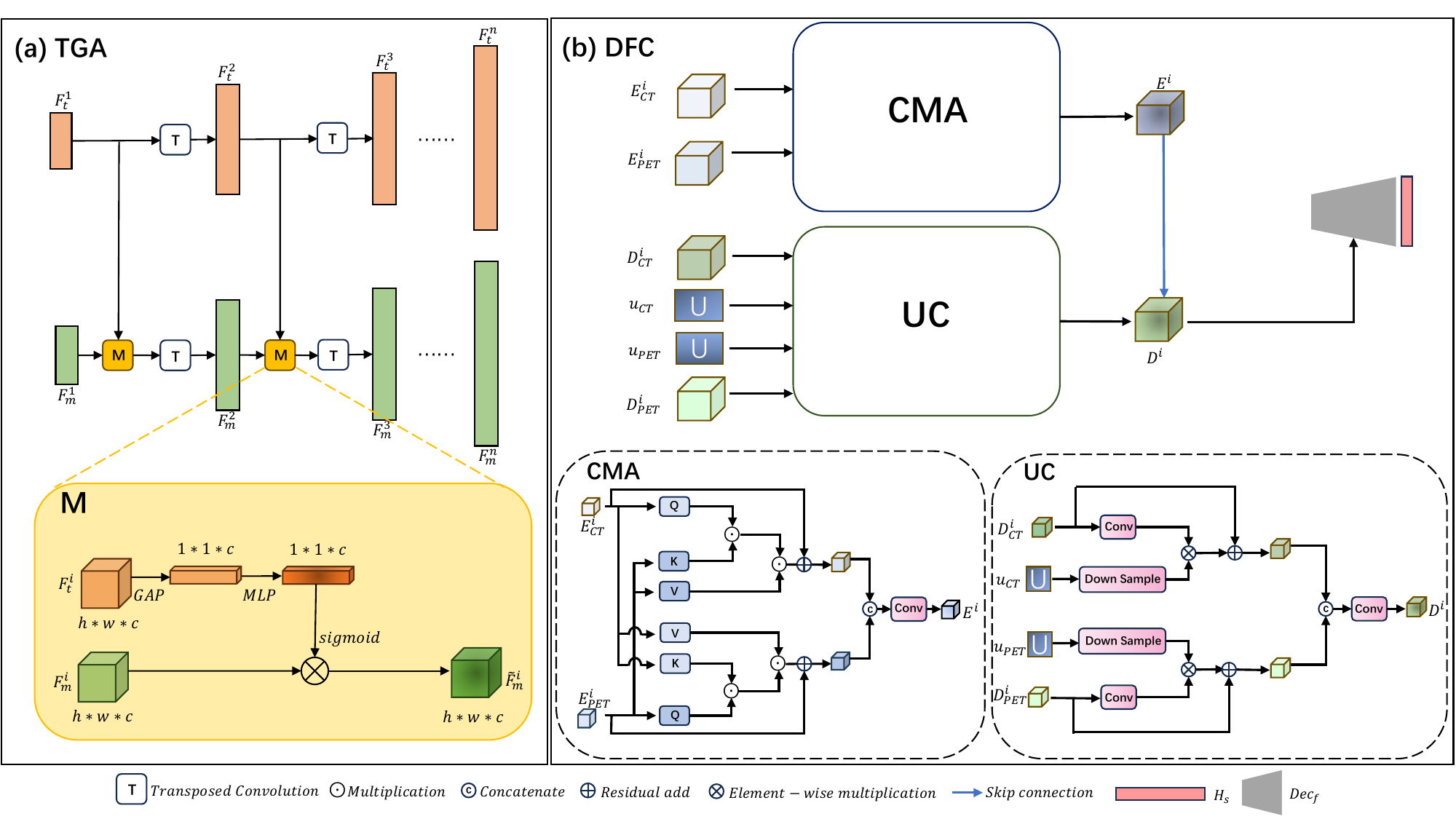}
    \caption{The architecture of (a) Tumor Guided Attention in CFL stage and (b) Dual-attention Feature Calibrating module in MTF stage.}
    \label{fig:TGA and DFC}
\end{figure*}

As illustrated in the top and bottom parts of Fig. \ref{fig:Overview}, our proposed MEFN encompasses two CFL branches with a structure mirrors  each other. Taking the CT branch as an example, given the input $X_{CT}$, the modality encoder $Enc^{CT}$ extracts high-level feature representations of the CT image. Subsequently, the modality decoder $Dec_{m}^{CT}$ transforms these high-dimensional features into modality-specific features and generates a synthetic PET image $\hat{X}_{PET}$. Concurrently, the tumor decoder $Dec_{t}^{CT}$ decouples tumor features from the high-dimensional features of the CT image to produce a synthetic tumor mask image $S_{Y}$. 
Notably, different to the segmentation task, $S_{Y}$ here represents the generated pseudo tumor mask image, rather than the probability prediction map of the tumor region. The task of modality discriminator $Dis_{m}^{PET}$ is to distinguish between $X_{PET}$ and $\hat{X}_{PET}$. To extract general tumor features across different modalities, a shared tumor discriminator $Dis_{t}$ is designed to differentiate between real tumor mask $Y$ and synthetic tumor mask $S_{Y}$. The architecture of all encoders, decoders and discriminator networks adheres to the framework established by Zhu et al. \citep{Zhu2017}. To further make the network focusing more on tumor features, we introduce the Tumor Guided attention (TGA) module. TGA is incorporated at every upsampling stage of $Dec_{m}$, where tumor features $F_{t}^{i}$ decoupled by $Dec_{t}$ are sent to Global Average Pooling (GAP) and a Multi-Layer Perceptron (MLP) in TGA module of $Dec_{m}$ (See Fig. \ref{fig:TGA and DFC}(a)) to learn channel attention. This channel attention is then employed to weight the modality features $F_{m}^{i}$ extracted by $Dec_{m}$ upon element-wise multiplication ($\otimes$). The detailed description of process flow of TGA is depicted in Fig. \ref{fig:TGA and DFC}(a). 

The entire CFL adopts adversarial training under the constraint of two adversarial losses for  generating fake PET (or CT) images and for generating fake tumor mask images, respectively. Taking the CT branch as an example, the above mentioned two losses denoted as $\mathcal{L}_{advm}^{CT}$ and $\mathcal{L}_{advt}^{CT}$ are defined as:
\begin{flalign}
    \begin{aligned}
    \mathcal{L}_{advm}^{C T}\left(Gen_m^{C T}, D i s_m^{C T}\right) &=-E_{X_{C T}, X_{P E T}}\left[\log D i s_m^{C T}\left(X_{C T}, X_{P E T}\right)\right] \\
    & -E_{X_{C T}}\left[\log \left(1-D i s_m^{C T}\left(X_{C T}, \hat{X}_{P E T}\right)\right)\right]
    \end{aligned} &&
\end{flalign}
\begin{flalign}
    \begin{aligned}
    \mathcal{L}_{advt}^{CT}\left(Gen_t^{CT}, Dis_t\right)&=-E_{X_{CT},Y}\left[\log Dis_t\left(X_{CT}, Y\right)\right]\\
    &-E_{X_{CT}}\left[\log \left(1-Dis_t\left(X_{CT}, S_Y\right)\right)\right]
    \end{aligned} &&
\end{flalign}
where $Gen_m^{CT}$ and $Gen_t^{CT}$ denote $Enc^{CT}$-$Dec_m^{CT}$  and $Enc^{CT}$-$Dec_t^{CT}$ encoder-decoder combinations, respectively. $X_{CT}$, $X_{PET}$, $\hat{X}_{PET}$, $Y$, $S_Y$ represent the real CT, real PET, fake PET, real tumor mask and predicted tumor mask, respectively. Considering the difference between the real target modality and the synthetic target modality, L1 loss is used to ensure that the generated target modality is close to the real target modality, thereby reducing blurriness. The L1 loss $\mathcal{L}_{l_1}^{CT}$ for CT branch can be defined as,
\begin{flalign}
    \begin{aligned}
        \mathcal{L}_{l_1}^{CT}\left(Gen_m\right)&=E_{X_{PET}, \hat X_{PET}}\left[\left\|X_{PET}-\hat X_{PET}\right\|_1\right]\\
    &+E_{Y, S_Y}\left[\left\|Y-S_Y\right\|_1\right]
    \end{aligned} &&
\end{flalign}
The same goes for the PET branch, so the loss function $\mathcal{L}_{CFL}$ of the entire CFL can be defined as,
\begin{flalign}
    \label{eq:CFL}
    \begin{aligned}
    \mathcal{L}_{CFL}&=\mathcal{L}_{advm}^{CT}+\mathcal{L}_{advt}^{CT}+\lambda_1 \mathcal{L}_{l_1}^{CT}\\
    &+\mathcal{L}_{advm}^{PET}+\mathcal{L}_{advt}^{PET}+\lambda_1 \mathcal{L}_{l_1}^{PET}
    \end{aligned} &&
\end{flalign}
where $\lambda_1$ is the hyperparameters used to balance the losses in Eq. (\ref{eq:CFL}). 
\label{sec:CFL}

\begin{figure}[t]
    \centering
    \includegraphics[width=0.9\columnwidth,trim=0 0 0 0,clip]{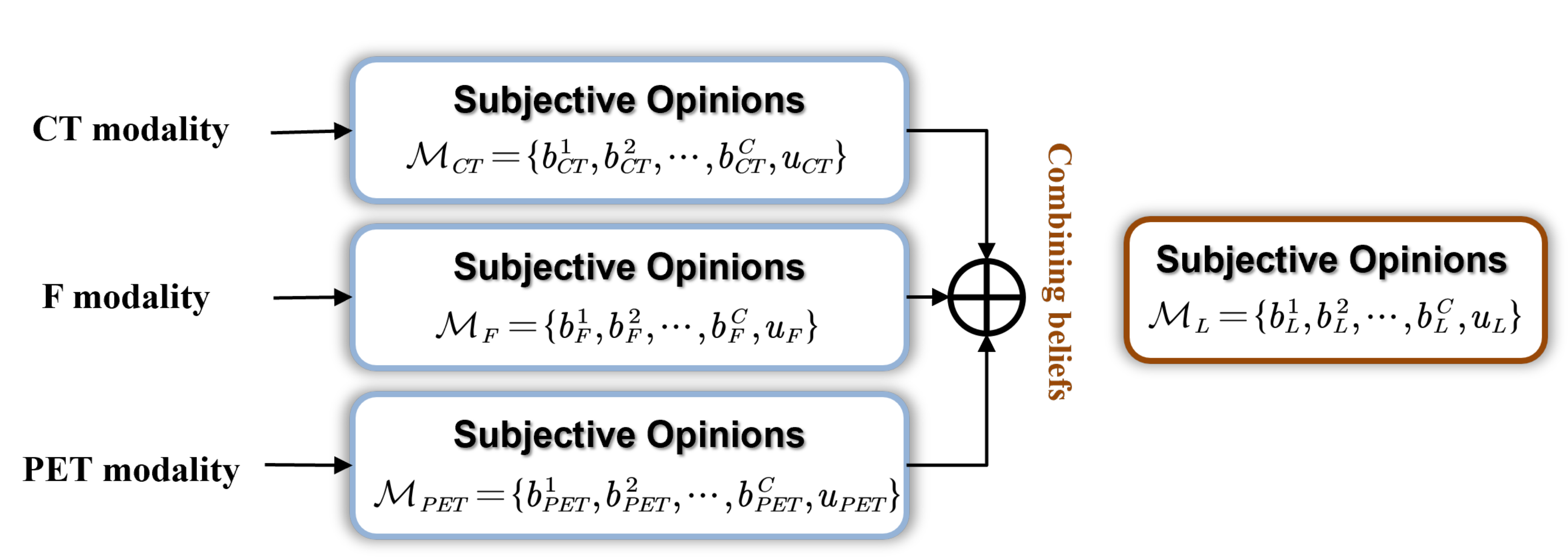}
    \caption{Schematic illustration of the Dempster-Shafer Theory-based Trustworthy Fusion.}
    \label{fig:DTF}
\end{figure}

\subsection{Segmentation Backbone and Multimodal Feature-levle Fusion}
 For convenient of  parameter sharing, the network architecture of the segmentation backbone is strikingly similar to the $Enc_m$-$Dec_m$ structure of the CFL, with the only modification being the final output layer. The CFL employs an output head denoted as $Ht$ ($Conv_{3,3}^1$+$Tahn$), while the segmentation backbone utilizes an output head denoted as $Hs$ ($Conv_{3,3}^{nc}$+$Softplus$). In pursuit of facilitating multi-modal feature fusion, we propose a novel fusion network. This network introduces a specifically designed fusion branch that leverages single-modality prediction outcomes and the modality's inherent uncertainties (details of which are delineated in section \ref{sec:Uncertainty Quantification and Dempster-Shafer Theory Based Trustworthy Fusion}) to guide the feature fusion stage. This approach not only allows for the seamless transfer of features learned from the CFL but also enables the acquisition of potent fusion features, which is instrumental for delineating the tumor regions required for segmentation.

Given modalities CT and PET, the multi-modal feature-level fusion network comprises two mono-modality segmentation br-anches of $Enc^{CT}$-$Dec_m^{CT}$ and $Enc^{PET}$-$Dec_m^{PET}$, as well as a multi-modal feature fusion branch of $Dec_f$. Herein, the mono-modality segmentation branch $Enc^{CT}$-$Dec_m^{CT}$ utilizes CT data as input to predict the segmentation mask $p_{CT}$ for the tumor region and its associated  uncertainty $u_{CT}$. Similarly, the mono-modality segmentation branch $Enc^{PET}$-$Dec_m^{PET}$ employs PET data as input, predicting the segmentation mask $p_{PET}$ for the tumor area and its corresponding  uncertainty $u_{PET}$.

To fuse effectively features from each modality, we introduce a novel Dual-attention Feature Calibrating (DFC) module as depicted in  Fig. \ref{fig:TGA and DFC}(b) which employs distinct fusion strategies for the encoder and decoder features of the two mono-modality branches. For encoder features, motivated by Vaswani et al. \citep{vaswani}, DFC utilizes Cross Modality Attention (CMA) as illustrated in the lower left part of Fig. \ref{fig:TGA and DFC}(b) which first merges  features $E_{CT}^i$ and $E_{PET}^i$ from different layers of the two mono-modality branches using query, key and value vectors and then  recovers the number of channels of the merged features through a $3\times3$ convolution to obtain the final fused feature $E^i$. Here, the query, key and value vectors are all derived from $1\times1$ convolutions. Upon  adaptive adjustments of these vectors, encoder features from each modality are selectively integrated, thereby reducing redundancy and improving fusion efficiency. For decoder features, DFC adopts an Uncertainty Calibrator (UC) as shown in the lower right part of Fig. \ref{fig:TGA and DFC}(b). The UC employs the predictive uncertainties $u_{CT}$ and $u_{PET}$ from the two mono-modality segmentation branches as calibrators for the decoder features, thereby refining the extracted semantic information. Specifically, for decoder features $D_{CT}^i$ and $D_{PET}^i$ of different layers, $U_{CT}$ and $U_{PET}$ are first down-sampled to match the dimensions of $D_{CT}^i$ and $D_{PET}^i$ whose results are then multiplied in element-wise manner by $1\times1$ convolutional results of $D_{CT}^i$ and $D_{PET}^i$, respectively. The resultant ones are  added residually to $D_{CT}^i$ and $D_{PET}^i$ respectively to alleviate degradation problem. Upon concatenating these two features in channel-wise manner and performing a $3\times3$ convolution,  the fused feature $D^i$ can be derived. In scenarios where the modalities exhibit areas of high uncertainty (typically at boundaries and conflicting features), UC directs the network's focus towards these areas of high uncertainty, thereby improve the network's precision in delineating boundary regions and reducing conflicts between modalities. Ultimately, $E^i$ and $D^i$ are connected in a manner similar to the skip connections in UNet and then fed into the multi-modal fusion branch $Dec_f$ to learn more potent fused features, culminating in the generation of a fused segmentation probability map $p_F$ and fused uncertainty $u_F$ for the tumor region.

Unlike conventional segmentation approaches, our segmentation backbone replaces Softmax in the traditional neural network classifier with Softplus to facilitate the quantification of uncertainty. Thus, the outcomes derived from the segmentation backbone should not be interpreted directly as the predictive segmentation results, but rather as evidence vectors $e$ representing the evidence observed by the model for classification. Consequently, parameters for the Dirichlet distribution can be obtained. For traditional neural network classifiers employing Softmax, the cross-entropy loss is commonly utilized:
\begin{flalign}
    \label{eq:cross-entropy loss}
    \begin{aligned}
    \mathcal{L}_{ce}=-\sum_{c=1}^C{y_{X}^{c}}\log\left(p_{X}^{c}\right)
    \end{aligned} &&
\end{flalign}
where $p_X^c$ and $y_X^c$ are the predicted probability and label that sample $X$ belongs to class $c$, respectively, $C$ is the number of classification. Due to the aforementioned difference between models with Softmax and those with Softplus,  the adjusted cross-entropy loss for our model can be derived by making simple modification with regard to the traditional cross-entropy loss in Eq. (\ref{eq:cross-entropy loss}) and can be formulated as:
\begin{flalign}
    \begin{aligned}
    \mathcal{L}_{ace}\left(\alpha_{X}\right)&=\int{\left[-\sum_{c=1}^{C}y_{X}^{c}\log\left(p_{X}^{c}\right)\right]}D\left(p_X \mid \alpha_X\right)dp_X\\
    &=\sum_{c=1}^C{y_{X}^{c}\left(\psi\left(S_X\right)-\psi\left(\alpha_{X}^{c}\right)\right)}
    \end{aligned} &&
\end{flalign}
where $\psi$ refers to the digamma function and $D\left(p_X \mid \alpha_X\right)$ is the Dirichlet probability density function over $p_X=[p_X^1,\dots,p_X^C]$ and can be expressed as:
    \begin{flalign}
    D\left(p_X \mid \alpha_X\right)= \begin{cases}\frac{1}{B\left(\alpha_{X}\right)}\prod_{c=1}^C{\left(p_{X}^{c}\right)^{\alpha_{X}^{c}-1}} & \text { for } p_X \in P_C \\ 0 & \text { otherwise }\end{cases}
\end{flalign}
Herein,   $\alpha_X=[\alpha_X^1,\dots,\alpha_X^C]$ (where $\alpha_X^c=e_X^c+1$ with $e$ being obtained from the segmentation backbone),  $p_X^c$ is  the projected probability and can be computed as:
\begin{flalign}
    p_X^c=1-\sum_{i\ne c}{b_X^i}
\end{flalign}
with $b_{X}^{i}=\small{\frac{e_X^i}{S_X}}=\small{\frac{\alpha_X^i-1}{S_X}}$ representing the belief mass, where $S_X=\sum\nolimits_{c=1}^C{\alpha_{X}^{c}}$ representing the Dirichlet strength, $B(\alpha_{X})=\int{\prod_{c=1}^C{(p_X^c)^{\alpha_X^c - 1}}dp_X}$ denotes the C-dimensional multinomial beta function and $P_C$ is the C-dimensional unit simplex given by:
\begin{flalign}
    P_C=\left\{p_X \mid 0 \leq p_X^1, \ldots, p_X^C \leq 1\right\}
\end{flalign}

 To ensure that incorrect labels generate less evidence, potentially reducing to zero, a KL divergence loss function is introduced as follows:
\begin{flalign}
    \begin{aligned}
    \mathcal{L}_{KL}\left(\alpha_{X}\right)&=\log\left(\frac{\varGamma\left(\sum\nolimits_{c=1}^C{\tilde{\alpha}_{X}^{c}}\right)}{\varGamma\left(\tilde{\alpha}_{X}^{c}\right)}\right)\\
    &+\sum\nolimits_{c=1}^C{\left(\tilde{\alpha}_{X}^{c}-1\right)}\left[\psi\left(\tilde{\alpha}_{X}^{c}\right)-\psi\left(\sum\nolimits_{c=1}^C{\tilde{\alpha}_{X}^{c}}\right)\right]
    \end{aligned} &&
\end{flalign}
where $\varGamma$ is the gamma function, $\tilde{\alpha}_{X}^{c}=y_{X}^{c}+\left(1-y_{X}^{c}\right)\odot\alpha_{X}^{c}$ ($\odot$ means element-wise multiplication) represents the adjustment parameters for the Dirichlet distribution employed to ensure that the evidence for the ground-truth class is not mistakenly considered to be zero. Moreover, due to the fact that the Dice score is a crucial metric for evaluating the performance of tumor segmentation, we employ the following soft Dice loss to optimize the network:
\begin{flalign}
    \begin{aligned}
    \mathcal{L}_{Dice}\left(p_{X}\right)=\sum_{c=1}^C{\left(1-\frac{2y_{X}^{c}p_{X}^{c}+smooth}{y_{X}^{c}+p_{X}^{c}+smooth}\right)}
    \end{aligned} &&
\end{flalign}
where $smooth$ represents the smooth coefficient which is empirically set to $1\times 10^{-5}$ in our work. Furthermore, to force the network focusing more on  prediction results for areas with greater uncertainty, we introduce the following Uncertainty Perceptual Loss:
\begin{flalign}
    \begin{aligned}
    \mathcal{L}_{UP}\left(p_X,u_X\right)=-u_X\sum_{c=1}^C{y_{X}^{c}\log\left(p_{X}^{c}\right)}
    \end{aligned} &&
\end{flalign}
where $u_X$ represents the uncertainty mass of the sample $X$ which will be defined in section \ref{sec:Uncertainty Quantification and Dempster-Shafer Theory Based Trustworthy Fusion}.

Given the substantial uncertainty in the model during the initial stages of training, to mitigate the impact of excessive uncertainty on model optimization, an annealing factor $(1-\beta_t)$ is employed, with $\beta_t=\beta_0e^{\left[-\left(\small{\frac{\ln\beta_0}{T}}\right)t\right]}$ while $T$ and $t$ represent the total number of epochs and the current epoch, respectively. Therefore, for the segmentation result $R_{CT}=\left\{\alpha_{CT},p_{CT},u_{CT}\right\}$ of CT, the overall segmentation loss function of our proposed network can be defined as follows:
\begin{flalign}
    \begin{aligned}
    \mathcal{L}_{seg}\left(R_{CT}\right)&=\left(1-\beta_t\right)\mathcal{L}_{ace}\left(\alpha_{CT}\right)+\beta_t \mathcal{L}_{UP}\left(p_{CT},u_{CT}\right)\\
    &+\mathcal{L}_{KL}\left(\alpha_{CT}\right)+\mathcal{L}_{Dice}\left(p_{CT}\right)
    \end{aligned} &&
    \label{eq:Lseg}
\end{flalign}

Similarly, $R_{PET}=\left\{\alpha_{PET},p_{PET},u_{PET}\right\}$ for the segmentation results of PET and $R_F=\left\{\alpha_{F},p_{F},u_{F}\right\}$ for the fused segmentation result, the segmentation losses $\mathcal{L}_{seg}(R_{PET})$ and $\mathcal{L}_{seg}(R_F)$ are derived in the same manner.
\label{sec:Segmentation Backbone and Cross-modality Feature-level Fusion}

\subsection{Uncertainty Quantification and Dempster-Shafer Theory Based Trustworthy Fusion}
Guided by Subjective Logic \citep{jsang2018} and based on the evidence gathered within the data, we may deduce an overall uncertainty (uncertainty masses) as well as probabilities for different categories (belief masses). Here, evidence denotes indicators collected from the input that support categorization, intimately linked to the density parameters of the Dirichlet distribution as discussed in section \ref{sec:Segmentation Backbone and Cross-modality Feature-level Fusion}. Based on the evidence vector $e$ output from the network, we can ascertain the parameters $\alpha_X$ of the Dirichlet distribution. Specifically, for a C-class problem, the subjective logic endeavors to allocate a belief mass to each category label and an overall uncertainty mass to the framework as a whole. Hence, for any modality $X$, the $C$ belief mass values $b_{X}^{c} (c=1,2,\ldots, C)$ and the overall uncertainty $u_X$ are non-negative and are cumulatively equating to one:
\begin{flalign}
    \begin{aligned}
    u_X+\sum_{c=1}^C{b_{X}^{c}}=1
    \end{aligned} &&
\end{flalign}
Thus, the overall uncertainty $u_X$ can be articulated as follows:
\begin{flalign}
    \begin{aligned}
    u_X=1-\sum_{c=1}^C{b_X^c}=\small{\frac{C}{S_X}}
    \end{aligned} &&
    \label{eq:uncertainty}
\end{flalign}
Eq. (\ref{eq:uncertainty}) implies that the more the evidence is observed for category $c$, the greater the probability will be assigned to $c$ and hence the less the overall uncertainty is. Conversely, the less evidence is observed in total, the greater the overall uncertainty will be. Belief assignments can be regarded as a form of subjective opinion.

 Intuitively, the  fused prediction of modalities $X_1$ and $X_2$ with high uncertainty (large $u_{X_1}$ and $u_{X_2}$) must possess low confidence (small $b^c$). Conversely, if the uncertainties of both modalities are low (small $u_{X_1}$ and $u_{X_2}$), the final prediction may exhibit higher confidence (large $b^c$). When only one modality demonstrates low uncertainty (only $u_{X_1}$ or $u_{X_2}$ is large), the final prediction relies solely on the modality with lower uncertainty. Hence, we employ the Dempster-Shafer Theory of Evidence (DST) to achieve our objectives. As shown in Fig. \ref{fig:DTF}, DST allows for the combination of evidence from different sources, which culminates in a belief (belief function) that considers all available evidence. Defining $M_{X_i}=\left\{\left\{b_{X_i}^c\right\}_{c=1}^C,u_{X_i}\right\}, (i=1,2)$ as the set of probability mass allocated for a single modality $X_i$, then  the joint mass  $M=\left\{\left\{b^c\right\}_{c=1}^C,u\right\}$ for modalities $X_1$ and $X_2$ can be defined as follows:
\begin{flalign}
    \begin{aligned}
    M=M_{X_1}\oplus M_{X_2}
    \end{aligned} &&
\end{flalign}
where $\oplus$ denotes the set operations defined as follows:
\begin{flalign}
    \begin{aligned}
    &b^c=\small{\frac{1}{1-Cof}\left(b_{X_1}^c b_{X_2}^c+b_{X_1}^c u_{X_2}+b_{X_2}^c u_{X_1}\right)}\\
    &u=\small{\frac{1}{1-Cof}}u_{X_1}u_{X_2}
    \end{aligned} &&
\end{flalign}
Herein, $Cof=\left(\sum\nolimits_{i\ne j}^{}{b_{X_1}^{i}}b_{X_2}^{j}+b_{X_1}^i u_{X_2}+b_{X_2}^j u_{X_1}\right) + u_1u_2$ represents the extent of confliction between two sets of masses and $\frac{1}{1-Cof}$ is employed for normalization purposes. In this work, to improve the efficiency of fusion,  we also incorporate the results of the DFC in addition to $CT$ and $PET$ modalities as a new modality $F$  into the decision-level fusion. Therefore, the final fusion function can be defined as follows:
\begin{flalign}
    \begin{aligned}
    M_L=M_{CT}\oplus M_{PET}\oplus M_{F}
    \end{aligned} &&
    \label{eq:fusion}
\end{flalign}
Upon obtaining the joint mass $M_L=\left\{\left\{b_L^c\right\}_{c=1}^C,u_L\right\}$, according to Eq. (\ref{eq:uncertainty}), the corresponding joint evidence $e_L^c$ and Dirichlet distribution parameters $\alpha_L^c$ are derived from multiple perspectives as follows:
\begin{flalign}
    \begin{aligned}
    S_L=\small{\frac{C}{u_L}},e_L^c=b_L^c\times S_L,\alpha_L^c=e_L^c+1\, \operatorname{and}\, p_L^c=b_L^c+u_L
    \end{aligned} &&
\end{flalign}

Following the aforementioned fusion rules, we can obtain the estimated multi-modal joint evidence 
$e_L=[e_L^1,\dots,e_L^C]$ and the corresponding parameters of the joint Dirichlet distribution 
$\alpha_L=[\alpha_L^1,\dots,\alpha_L^C]$ as well as the final prediction of the multimodal Trustworthy fusion 
$p_L=[p_L^1,\dots,p_L^C]$ and the overall uncertainty $u_L$. For deep supervision of network training, the final segmentation result $R_L=\left\{\alpha_{L},p_{L},u_{L}\right\}$ from decision-level fusion can be obtained in a similar manner to the previously defined Eq. (\ref{eq:Lseg}) for the loss $\mathcal{L}_{seg}(R_L)$.
\label{sec:Uncertainty Quantification and Dempster-Shafer Theory Based Trustworthy Fusion}
\label{sec:Methodology}

\section{Experiments and discussions}
\subsection{Datasets and perprocessing}
The proposed segmentation framework is evaluated using two public PET/CT datasets: the AutoPET dataset from the 25th MICCAI Challenge on whole-body PET/CT segmentation \citep{gatidis2022whole} and the Hecktor dataset from the 23rd MICCAI Challenge on head and neck tumor segmentation in PET/CT \citep{Andrearczyk2022head}.

The AutoPET dataset comprises 900 histologically confirm-ed malignant melanoma, lymphoma or lung cancer patients from two large medical centers in Germany who undergo PET/CT examinations. Two experienced radiologists manually delineate all tumor regions, tasked with whole-body tumor segmentation.

The Hecktor dataset consists of 882 patients with primary tumors and with gross tumor volumes of the head and neck from four different centers. All patients undergo FDG-PET/CT imaging scans within 18 days before treatment (range: 6-66). It contains two segmentation tasks, one for the primary tumor (GTVp) in the head and neck and another for the involved lymph nodes (GTVn).

Given data from multi-centers, we normalize the CT and PET images. The CT volumes are clipped to [-1024, 1024] Hounsfield units and then mapped to [-1, 1]. We normalize the PET using the Z-score. As the segmentation is performed in 2D, we sliced the 3D datasets. After which we resize them to 256 $\times$ 256 and then add 25$\%$ negative samples to each dataset for data balance. In total, our experiments comprise 32,158 pairs of PET/CT slices from the Hecktor dataset and 29,065 pairs of PET/CT slices from the AutoPET dataset. We conducte 5-fold cross-validation, where each dataset is randomly partitioned into 5 mutually exclusive subsets based on different patients, so as to ensure that different subsets do not contain slices of the same patients. 

\label{sec:Datasets and perprocessing}
\begin{figure*}[t]
    \centering
    \includegraphics[width=\textwidth,trim=0 0 0 0,clip]{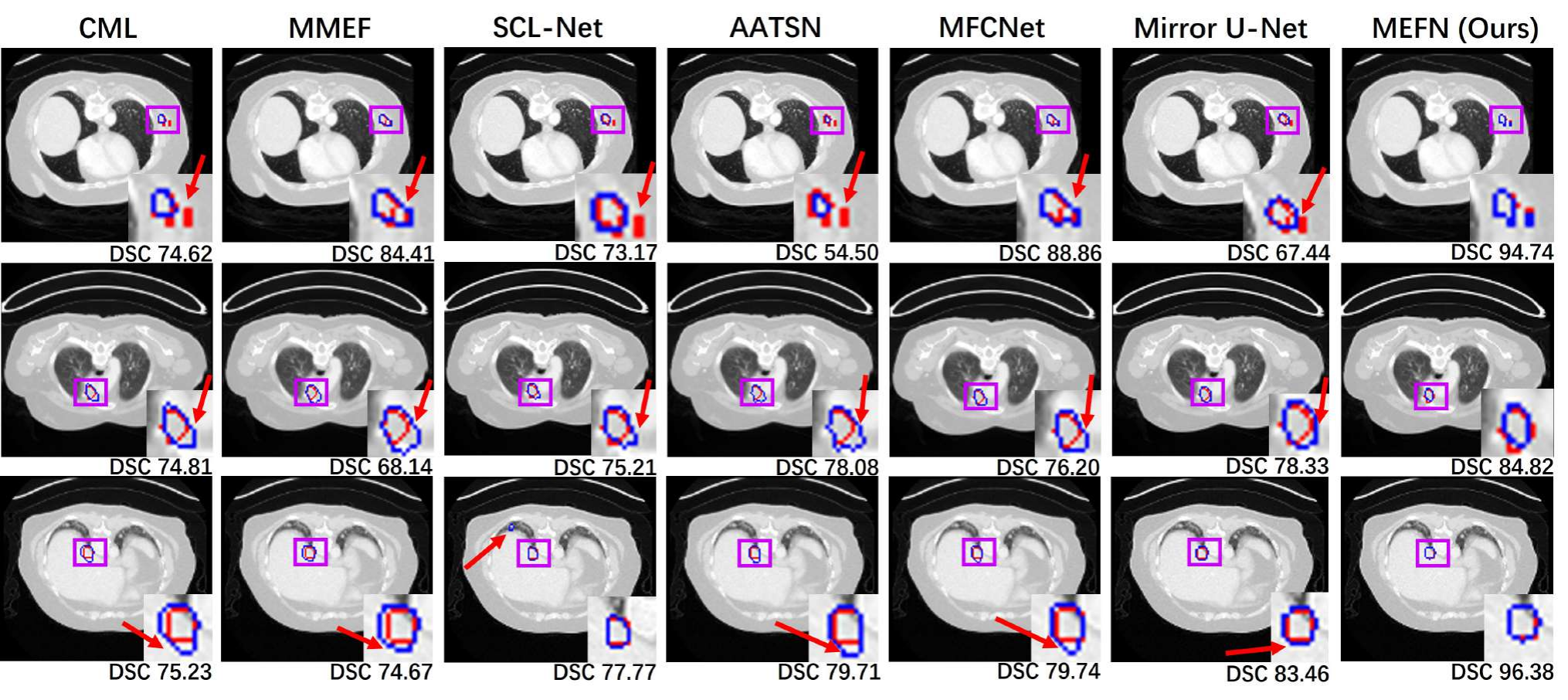}
    \caption{\justifying Segmentation results (blue contour) from different multi-modal segmentation methods of CT sample images for three patients from the AutoPET dataset.  The ground truth is displayed in red contour. Each row represents segmentation results for the same patient. The bottom right corner of each image displays the enlarged version of the region indicated by the purple box. Red arrows highlight the mis-segmented regions.}
    \label{fig:contrasted exp-autopet}
\end{figure*}

\begin{figure*}[t]
    \centering
    \includegraphics[width=\textwidth,trim=0 0 0 0,clip]{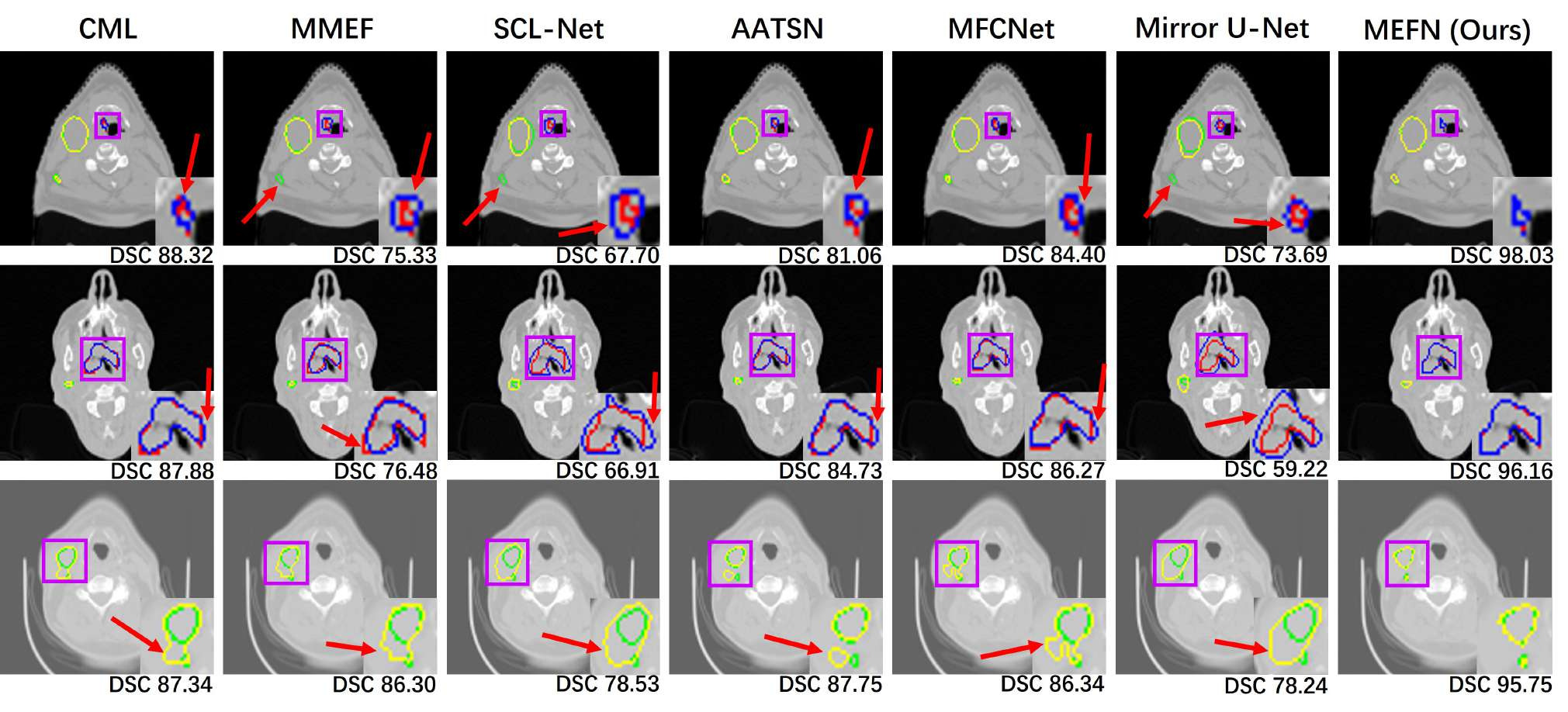}
    \caption{\justifying Segmentation results from different multi-modal segmentation methods of CT sample images for different patients on the Hecktor dataset. Blue and yellow contours represent the segmentation results of the GTVp and GTVn tasks, respectively. Red and green contours represent the ground truth of the GTVp and GTVn tasks, respectively. Each row represents the segmentation results of the same patient. The bottom right corner of each image displays the enlarged version of region indicated by the purple box. Red arrows highlight the mis-segmented regions.}
    \label{fig:contrasted exp-hecktor}
\end{figure*}

\subsection{Implementation details and metrics}\label{sec:metric}

The proposed approach is implemented using Python and the PyTorch \citep{paszke2019pytorch} library on a PC equipped with an NVIDIA GTX 3090 GPU (24 GB memory). The training is conducted using the Adam optimizer with a momentum of 0.99. Based on empirical observations, the learning rate is set to $1 \times 10^{-4}$ and the number of epoch is set to 100. The parameter $\lambda_1$ in Eq. (\ref{eq:CFL}) is determined upon grid search on candidate values \{1, 50, 100, 150, 200\}. The value of 100  with the highest validation accuracy is chosen for our subsequent experiments.
The segmentation model's performance is evaluated in terms of the Dice similarity coefficient (DSC), Jaccard, 95$\%$ Hausdorff Distance (HD95), sensitivity (Sens) and precision (Pre).
\label{sec:Implementation details and metrics}
\begin{table*}[t]
    \centering
    \caption{Segmentation performance of MEFN and other state-of-the-art methods on AutoPET and Hecktor datasets (Mean $\pm$ std). The best results are highlighted in bold and the second-best ones are underlined.}
    \label{tab:comparsion exp}
    \resizebox{0.95\textwidth}{!}{%
    \begin{tabular}{@{}lllllll@{}}
    \toprule
    Dataset                                      & Methods & DSC(\%)       & Jaccard(\%)   & HD95(mm)       & Sens(\%)      & Pre(\%)       \\
    \midrule
    \multirow{6}{*}{AutoPET}
    & CML (2021) \citep{zhang2021}     & 78.88 $\pm$ 27.18          & \underline{71.25 $\pm$ 27.83}          & \underline{3.62 $\pm$ 12.26}          & 84.00 $\pm$ 21.29          & 83.73 $\pm$ 25.41          \\
    & MMEF (2021) \citep{huang2021}     & 78.70 $\pm$ 27.81          & 71.24 $\pm$ 28.26          & 3.91 $\pm$ 9.03          & \underline{85.46 $\pm$ 21.25}          & 82.27 $\pm$ 25.90          \\
    & SCL-Net (2023) \citep{wang2023SCL-Net}   & 71.35 $\pm$ 32.51          & 63.46 $\pm$ 32.25          & 5.32 $\pm$ 11.05          & 76.99 $\pm$ 30.52          & 80.30 $\pm$ 26.60          \\
    & AATSN (2023) \citep{ahmad2023}   & 69.28 $\pm$ 32.94          & 61.10 $\pm$ 32.76          & 7.29 $\pm$ 14.76          & 75.11 $\pm$ 32.77          & 80.12 $\pm$ 26.14          \\
    & MFCNet (2023) \citep{wang2023}  & \underline{79.35 $\pm$ 21.63}          & 70.54 $\pm$ 27.16          & 4.08 $\pm$ 9.26          & 84.62 $\pm$ 22.86          & \underline{84.41 $\pm$ 26.18}          \\
    &
    Mirror U-Net (2023) \citep{marinov2023}  & 75.80 $\pm$ 30.08           & 68.22 $\pm$ 30.24          & 4.32 $\pm$ 10.10               & 81.06 $\pm$ 27.59              & 82.83 $\pm$ 24.03         \\
    & MEFN (Ours)    & \pmb{82.45 $\pm$ 13.21} & \pmb{75.32 $\pm$ 27.26} & \pmb{2.96 $\pm$ 7.75} & \pmb{86.23 $\pm$ 20.55} & \pmb{88.55 $\pm$ 23.63} \\
    \midrule
    \multicolumn{1}{c}{\multirow{6}{*}{Hecktor}} & CML (2021) \citep{zhang2021}     & \underline{80.12 $\pm$ 25.32}  & 63.27 $\pm$ 27.59   & 4.26 $\pm$ 7.21    & 76.73 $\pm$ 33.74   & 86.32 $\pm$ 19.61       \\
    \multicolumn{1}{c}{}                         & MMEF (2021) \citep{huang2021}   & 78.11 $\pm$ 23.95          & 61.11 $\pm$ 28.63          & \underline{4.16 $\pm$ 7.02}          & 69.09 $\pm$ 32.14          & \underline{86.80 $\pm$ 17.48}         \\
    \multicolumn{1}{c}{}                         & SCL-Net (2023) \citep{wang2023SCL-Net}   & 72.42 $\pm$ 24.03          & 62.84 $\pm$ 27.46          & 9.80 $\pm$ 11.82          & 71.84 $\pm$ 34.62          & 75.98 $\pm$ 28.63          \\
    \multicolumn{1}{c}{}                         & AATSN (2023) \citep{ahmad2023}   & 76.73 $\pm$ 25.21          & \underline{65.13 $\pm$ 26.91}          & 12.24 $\pm$ 20.85           & \pmb{78.31 $\pm$ 27.58}          & 79.43 $\pm$ 23.13          \\
    \multicolumn{1}{c}{}                         & MFCNet (2023) \citep{wang2023}  & 74.14 $\pm$ 27.72          & 62.96 $\pm$ 22.41          & 6.41 $\pm$ 7.01           & 77.20 $\pm$ 27.13          & 76.12 $\pm$ 28.85          \\
    \multicolumn{1}{c}{}                         & Mirror U-Net (2023) \citep{marinov2023}    & 76.07 $\pm$ 23.49              & 61.56 $\pm$ 26.99              &  4.55 $\pm$ 9.91              & 71.50 $\pm$ 33.27              & 73.56 $\pm$ 25.16             \\
    \multicolumn{1}{c}{}                         & MEFN (Ours)    & \pmb{83.35 $\pm$ 23.09} & \pmb{71.73 $\pm$ 28.02} & \pmb{3.27 $\pm$ 7.21}  & \underline{78.03 $\pm$ 29.47} & \pmb{90.06 $\pm$ 16.07} \\ \bottomrule
    \end{tabular}%
    }
\end{table*}

\subsection{Comparisons with the state-of-the-art methods}

We conducted  comparative segmentation experiments on both the AutoPET and Hecktor datasets. Six most recent multi-modal tumor segmentation methods, namely CML \citep{zhang2021}, MMEF \citep{huang2021}, SCL-Net \citep{wang2023SCL-Net}, AATSN \citep{ahmad2023}, MFCNet \citep{wang2023} and Mirror U-Net \citep{marinov2023}, are compared, whose results are shown in Table \ref{tab:comparsion exp}.

On the AutoPET dataset, CML \citep{zhang2021} achieved an improved performance with a DSC score of 78.88$\%$ and Jaccard score of 71.25$\%$. This can be attributed to the introduction of the cross-modality Feature Transition Module to allevate the impact of domain gap between PET and CT before segmentation. MFCNet \citep{wang2023} further improves the DSC score to 79.35$\%$ by integrating a calibration network but sacrifices the accuracy of tumor edges as indicated by the increased HD95 compared to CML \citep{zhang2021} (by 0.46mm). Upon utilizing the Dempster-Shafer Theory for feature fusion strategy, MMEF \citep{huang2021}  achieves the second best sensitivity of 85.46$\%$ among all methods. It is worth noting that our MEFN outperforms all other models in terms of metrics. Specifically, the DSC score is improved by 3.10$\%$ compared to the second-best model, MFCNet \citep{wang2023}.
On the Hecktor dataset, AATSN \citep{ahmad2023} performs well due to its lightweight attention mechanism, achieving a sensitivity score of 78.31$\%$ and ranking the second in Jaccard (65.13$\%$). Additionally, CML \citep{zhang2021} achieves high DSC score and precision (80.12$\%$ and 86.32$\%$, respectively) due to its unique multi-modal feature fusion mechanism. Notably, our MEFN outperforms other segmentation networks. Specifically, as compared with the second-best network, our MEFN improves the DSC score by 3.23$\%$.

Besides the aforementioned quantitative comparisons among different methods, visual assessments and  comparisons are also conducted between MEFN and other state-of-the-art multi-modal tumor segmentation methods.  Fig. \ref{fig:contrasted exp-autopet} gives the segmented results from different methods for three different CT sample images of patients from AutoPET dataset where blue contour gives the predicted tumor region while red  contour gives the ground truth.  Fig. \ref{fig:contrasted exp-hecktor} gives the segmentation results for three CT sample images of three patients from the Hecktor dataset  where blue and yellow contours give the predicted  regions for GTVp and GTVn tumors  while red and green contours give the ground truths for these two kinds of tumors, respectively. Tumor regions marked with purple boxes are enlarged and placed on the right bottom of each sample image. Red arrows point out the mis-segmented tumor regions in each sample image. From Fig. \ref{fig:contrasted exp-autopet}, it can be seen that for small tumors throughout the body, such as in the first patient, CML \citep{zhang2021}, SCL-Net \citep{wang2023SCL-Net}, Mirror U-Net \citep{marinov2023} and AATSN \citep{ahmad2023} lead to under segmented results, while MMEF \citep{huang2021} and MFCNet \citep{wang2023} tend to merge two separate tumors into one large tumor. Additionally, for the second and third patients, the segmentation results of other models also differ in shape from the ground truth and exhibit many redundant edges. In contrast, our MEFN not only effectively segments all small tumors but also achieves the highest accuracy in fitting the tumor edges. 

The performance improvements of the MEFN algorithm stem from three key enhancements. First, MEFN introduces a novel modality translation stage, CFL, which aligns features across modalities and learns more robust feature representations. Unlike traditional approaches, which struggle to effectively focus on tumor regions, MEFN incorporates TGA, a targeted strategy that enhances the network's attention to these critical areas. Second, MEFN employs an advanced attention mechanism for feature fusion via the DFC module. This approach outperforms conventional concatenation-based methods \citep{huang2021} and single attention mechanisms \citep{ahmad2023}. Within DFC, the MCA in the encoder and UC in the decoder leverage mutual attention and uncertainty calibration to minimize redundant and conflicting features, significantly improving segmentation efficiency. Finally, the DTF module optimizes decision-level fusion by assigning weights based on uncertainty. Results with higher uncertainty are given lower weights, ensuring more reliable segmentation outcomes.

\label{sec:Comparisons with the state-of-the-art methods}

\subsection{Ablative study}
\subsubsection{Effectiveness of each component}
\begin{table*}[t]
    \caption{Effectiveness of each component on AutoPET and Hecktor datasets (Mean $\pm$ std). The best results are highlighted in bold and the second-best ones are underlined.}
    \centering
    \label{tab:ablation exp}
    \resizebox{\textwidth}{!}{%
    \begin{tabular}{@{}llllllllll@{}}
    \toprule
    Dataset    & Methods  & DTF & TGA & DFC  & DSC(\%)       & Jaccard(\%)   & HD95(mm)       & Sens(\%)      & Pre(\%) \\
    \midrule
    \multirow{5}{*}{AutoPET}
    & Baseline     & -   & -    & -       & 79.91 $\pm$ 27.72          & 73.86 $\pm$ 31.51          & 3.50 $\pm$ 10.29          & 80.26 $\pm$ 24.27          & 87.31 $\pm$ 23.82          \\
    & Baseline+DTF   & $\checkmark$   & -    & -     & 81.63 $\pm$ 25.17          & 74.43 $\pm$ 26.31          & 4.68 $\pm$ 10.81          & 83.31 $\pm$ 22.41          & \underline{88.18 $\pm$ 20.59}          \\
    & Baseline+DTF+TGA & $\checkmark$   & $\checkmark$    & -  & 82.29 $\pm$ 25.46          & 75.22 $\pm$ 26.45          & 3.05 $\pm$ 7.91          & \underline{86.17 $\pm$ 19.95}          & 86.29 $\pm$ 22.79          \\
    & Baseline+DTF+DFC  &$\checkmark$   & -    & $\checkmark$ & \underline{82.43 $\pm$ 24.29}          & \underline{75.28 $\pm$ 25.64}          & \underline{3.05 $\pm$ 7.91}          & 85.07 $\pm$ 21.16    & 87.33 $\pm$ 20.71          \\
    & MEFN (Ours)    & $\checkmark$   & $\checkmark$    & $\checkmark$      & \pmb{82.45 $\pm$ 13.21} & \pmb{75.32 $\pm$ 27.26} & \pmb{2.96 $\pm$ 7.75} & \pmb{86.23 $\pm$ 20.55}          & \pmb{88.55 $\pm$ 23.63} \\ \midrule
    \multirow{5}{*}{Hecktor}
    & Baseline     & -   & -    & -      & 81.42 $\pm$ 31.21          & 69.07 $\pm$ 32.24          & 3.68 $\pm$ 7.27          & 72.32 $\pm$ 27.75          & 89.31 $\pm$ 32.65          \\
    & Baseline+DTF   & $\checkmark$   & -    & -    & 81.63 $\pm$ 25.71          & 71.22 $\pm$ 29.30          & 3.52 $\pm$ 7.09          & \underline{77.19 $\pm$ 30.92}          & 89.43 $\pm$ 15.84          \\
    & Baseline+DTF+TGA & $\checkmark$   & $\checkmark$    & - & 82.45 $\pm$ 23.02          & 69.49 $\pm$ 28.60          & 3.62 $\pm$ 7.32          & 74.38 $\pm$ 30.03          & \underline{89.67 $\pm$ 15.63}          \\
    & Baseline+DTF+DFC & $\checkmark$   & -    & $\checkmark$ & \underline{83.13 $\pm$ 23.55}          & \underline{71.44 $\pm$ 26.31}          & \underline{3.34 $\pm$ 8.29}          & 73.31 $\pm$ 22.41          & 88.18 $\pm$ 20.59          \\
    & MEFN (Ours)   & $\checkmark$   & $\checkmark$    & $\checkmark$      & \pmb{83.35 $\pm$ 23.09} & \pmb{71.73 $\pm$ 28.02} & \pmb{3.27 $\pm$ 7.21} & \pmb{78.03 $\pm$ 29.47} & \pmb{90.06 $\pm$ 16.07} \\ \bottomrule
    \end{tabular}%
    }
\end{table*}
To better perform multi-modal segmentation, DTF, TGA and DFC modules are incorporated in our segmentation model. To evaluate the effectiveness of each component, ablation experiments are conducted on the AutoPET and Hecktor datasets, with the results presented in Table \ref{tab:ablation exp}. The baseline network  for comparison is the UNet without the DTF, TGA and DFC modules. In Baseline, PET features and CT features are fused simply by concatenation and input to $Dec_f$ to obtain the final result. It is worth noting that baseline still contains CFL to align features across different modalities and the outputs of CT branch and PET branch are used for deep supervision to constrain the fusion result. Baseline+DTF incorporates the DTF module into the Baseline model. Baseline+DTF+TGA further adds TGA, while Baseline+DTF+DFC integrates DFC into Baseline+DTF. Finally, upon incorporating all three components into the Baseline we obtain our proposed MEFN. Notably, as shown in Table \ref{tab:ablation exp}, our baseline surpasses several state-of-the-art methods. This may be attributed to our use of CFL to learn more robust feature representations and the application of results from the CT and PET branches as deep supervision. On the AutoPET dataset, the result of Baseline+DTF is significantly superior to that of the Baseline  model (81.63$\%$ vs. 79.91$\%$ in DSC), which indicates that the DTF module effectively utilizes the uncertainties of PET result, CT result and fusion result to achieve more reasonable weight allocation in the decision-level fusion. After incorporating TGA, the network focuses more on the tumor regions, making the model achieve a noticeable improvement in HD95 and sensitivity. With the introduction of DFC, the result is improved compared to Baseline+DTF (82.43$\%$ vs. 81.63$\%$ in DSC). Such improvement may stem from the mutual attention mechanism of CMA and the uncertainty calibration mechanism of UC which effectively reduce the redundant features and conflict features in the training process of the network. After incorporating all components, further improvements can be observed, which indicates that all modules contribute corporatively to the proposed method.
\label{sec:Ablation of each component}

\subsubsection{Effectiveness of Uncertainty Calibrator}
\begin{figure*}[t!]
    \centering
    \includegraphics[width=0.8\textwidth,trim=0 0 0 0,clip]{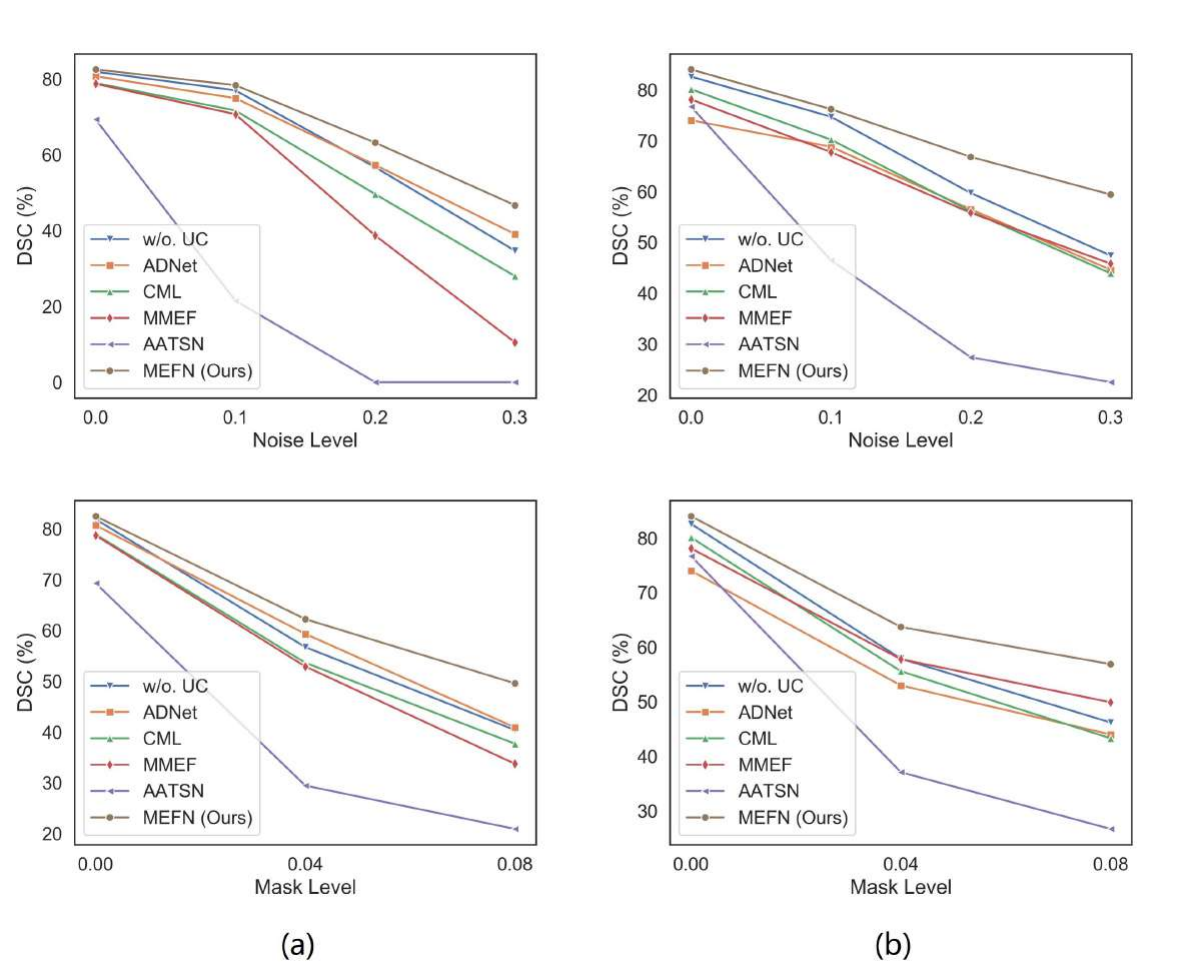}
    \caption{\justifying The performance of MEFN, MEFN without UC (w/o.UC) and other multi-modal segmentation models on two datasets with different levels of Gaussian noise perturbations and patch-size random masking: (a) AutoPET Dataset, (b) Hecktor Dataset.}
    \label{fig:UC exp}
\end{figure*}
In this work, the Uncertainty Calibrator (UC) is proposed to achieve semantic correction of decoder features during feature fusion. To validate the effectiveness of the proposed method, we introduced perturbations of Gaussian noise at different levels (with noise variance $\sigma^2=0.1,0.2,0.3$) and patch-size random masking (mask ratio $\gamma=0.04,0.08$) into the AutoPET and Hecktor datasets, respectively. For patch-size random masking, we use 3$\times$3 black boxes to cover the original input image with a uniform distribution, and the mask ratio $\gamma$ represents the proportion of black boxes to the total number of pixels.  We compare the performance of our MEFN, MEFN without UC (referred to as w/o. UC) and other models on out-of-distribution data (with large uncertainty), whose results are shown in Fig. \ref{fig:UC exp}.  From this figure,  it can be observed that as the perturbation level increases, the performance of other models deteriorate drastically, while our proposed MEFN performs much more robust to random noise and random masking. We notice that after removing UC, the performance degradation rate of our model without UC is comparable to CML \citep{zhang2021} (blue line and green line), which indicates that the model loses the ability to calibrate semantic features in large uncertainty regions after removing UC from our MEFN model.
\label{sec:Analysis on Uncertainty Calibrator}

\subsubsection{Effectiveness of Uncertainty Perceptual Loss}
\begin{table*}[t]
    \centering
    \caption{The performance of MEFN trained under different loss functions on AutoPET and Hecktor datasets (Mean $\pm$ std). The best results are highlighted in bold and the second-best ones are underlined.}
    \label{tab:loss exp}
    \resizebox{0.9\textwidth}{!}{%
    \begin{tabular}{@{}lllllll@{}}
    \toprule
    Dataset  & Methods  & DSC(\%)  & Jaccard(\%)  & HD95(mm)  & Sens(\%)   & Pre(\%)  \\
    \midrule
    \multirow{5}{*}{AutoPET} & $\mathcal{L}_{BS}$
    & 77.92 $\pm$ 31.45    & 71.56 $\pm$ 31.25   & 3.31 $\pm$ 7.55  & 83.96 $\pm$ 24.28   & 84.94 $\pm$ 26.28            \\
    & $\mathcal{L}_{BS} + \mathcal{L}_{U}$ \citep{huang2021evidential} & \underline{78.65 $\pm$ 31.27}  & \underline{75.19 $\pm$ 31.76}          & \underline{3.06 $\pm$ 7.83}           & \underline{84.94 $\pm$ 31.42}          & \underline{87.76 $\pm$ 17.82}          \\
    & $\mathcal{L}_{BS} + \mathcal{L}_{CU}$ \citep{zou2023evidencecap}  & 77.93 $\pm$ 30.14 & 71.23 $\pm$ 30.69          & 3.49 $\pm$ 7.91 & 82.92 $\pm$ 25.17          & 85.26 $\pm$ 24.27         \\
    & $\mathcal{L}_{BS} + \mathcal{L}_{CU} (dis)$ \citep{zou2023evidencecap}  & 78.46 $\pm$ 30.26 & 72.06 $\pm$ 31.72          & 3.46 $\pm$ 7.90          & 83.73 $\pm$ 26.42          & 85.46 $\pm$ 25.91 \\
    & $\mathcal{L}_{BS} + \mathcal{L}_{UP}$ (Ours) & \pmb{82.45 $\pm$ 13.21}            & \pmb{75.32 $\pm$ 27.26} & \pmb{2.96 $\pm$ 7.75}          & \pmb{86.23 $\pm$ 20.55} & \pmb{88.55 $\pm$ 23.63}          \\ \midrule
    \multirow{5}{*}{Hecktor} & $\mathcal{L}_{BS}$           & 81.46 $\pm$ 26.17                     & 65.65 $\pm$ 33.23          & 4.39 $\pm$ 9.14          & 69.47 $\pm$ 35.09          & 91.06 $\pm$ 19.40          \\
    & $\mathcal{L}_{BS} + \mathcal{L}_{U}$ \citep{huang2021evidential}        & \underline{83.03 $\pm$ 23.17}                     & \underline{70.40 $\pm$ 34.21}          & \underline{3.20 $\pm$ 7.83}          & 72.13 $\pm$ 34.32          & \underline{92.19 $\pm$ 19.17}          \\
    & $\mathcal{L}_{BS} + \mathcal{L}_{CU}$ \citep{zou2023evidencecap}       & 82.44 $\pm$ 25.71                     & 69.76 $\pm$ 33.20          & \pmb{3.19 $\pm$ 7.92}          & \underline{72.96 $\pm$ 34.61}          & \pmb{92.92 $\pm$ 16.12}          \\
    & $\mathcal{L}_{BS} + \mathcal{L}_{CU} (dis)$ \citep{zou2023evidencecap}  & 79.87 $\pm$ 31.91 & 51.61 $\pm$ 42.85            & 4.64 $\pm$ 8.35          & 52.46 $\pm$ 43.17          & 91.80 $\pm$ 19.46          \\
    & $\mathcal{L}_{BS} + \mathcal{L}_{UP}$ (Ours) & \pmb{83.35 $\pm$ 23.09}            & \pmb{71.73 $\pm$ 28.02} & 3.27 $\pm$ 7.21 & \pmb{78.03 $\pm$ 29.47} & 90.06 $\pm$ 16.07 \\ \bottomrule
    \end{tabular}%
    }
\end{table*}

\begin{table}[t]
    \centering
    \caption{\justifying Final DSC ($\%$) with different initialization schemes (Mean $\pm$ std). The best results are highlighted in bold and the second-best ones are underlined.}
    \label{tab:init exp}
    \resizebox{0.4\columnwidth}{!}{%
    \begin{tabular}{@{}lll@{}}
    \toprule
    Type       & AutoPET & Hecktor \\ \midrule
    Random     & \underline{77.87 $\pm$ 27.45}       & 82.28 $\pm$ 15.19    \\
    Xavier \citep{Xavier}     & 75.69 $\pm$ 23.17      & 82.09 $\pm$ 28.71    \\
    Kaiming \citep{he2015}    & 77.45 $\pm$ 29.17        & 80.64 $\pm$ 21.24    \\
    Orthogonal \citep{saxe2013} &  45.14 $\pm$ 47.17       & \underline{82.44 $\pm$ 26.90}    \\
    CFL (Ours) & \pmb{82.45 $\pm$ 13.21}    & \pmb{83.35 $\pm$ 23.09}    \\ \bottomrule
    \end{tabular}%
    }
\end{table}

\begin{figure}[t!]
    \centering
    \includegraphics[width=0.6\columnwidth,trim=0 0 0 0,clip]{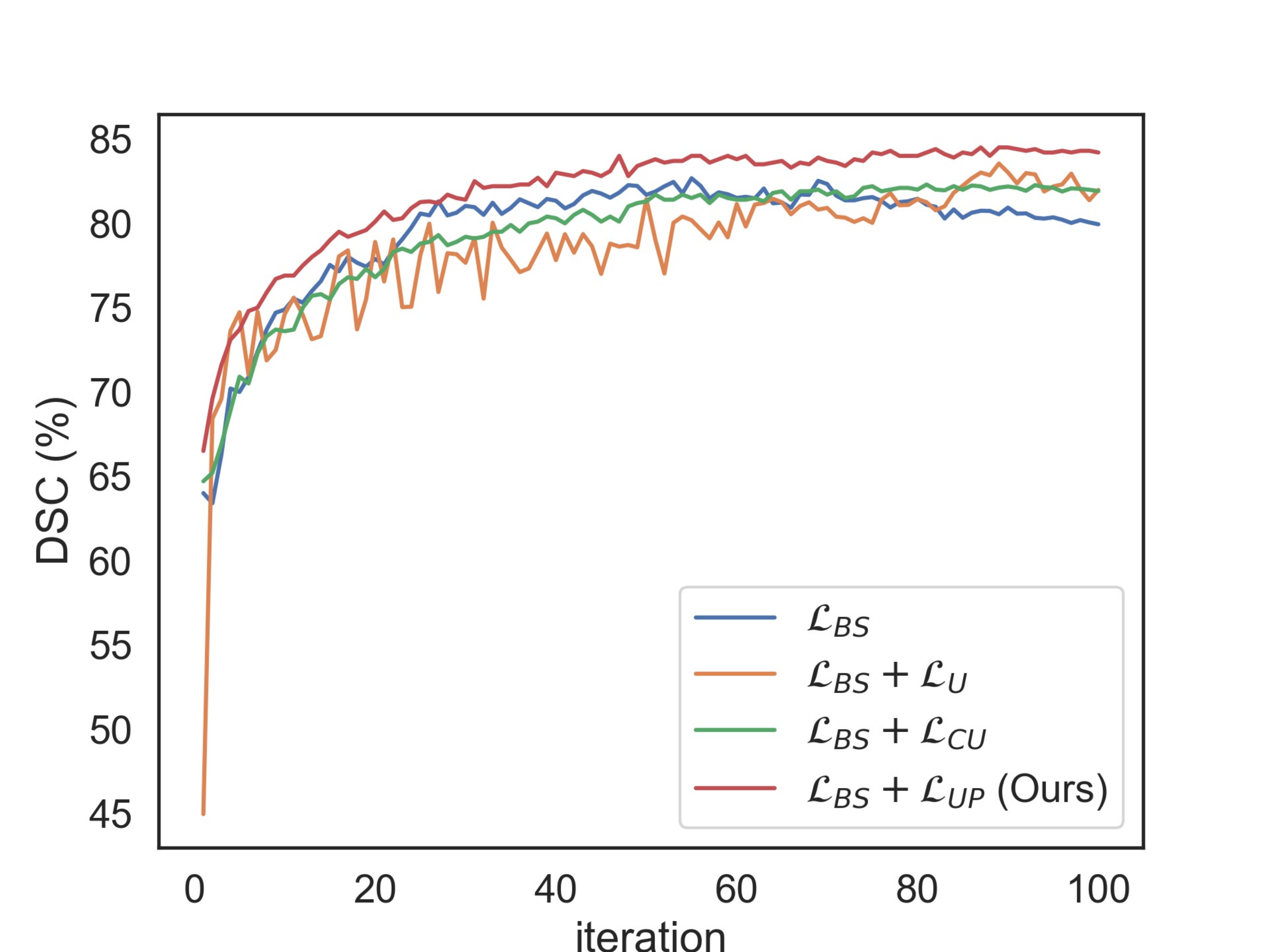}
    \caption{The DSC of MEFN during its training process under different loss functions on the test set of Hecktor dataset.}
    \label{fig:loss exp}
\end{figure}
In our work, the Uncertainty Perceptual Loss ($\mathcal{L}_{UP}$) is proposed to provide supervision for uncertainty. To validate the effectiveness of $\mathcal{L}_{UP}$ in model training, we compare performance of three models trained respectively under the combination of the baseline loss and other three different loss functions used for supervising uncertainty: uncertainty loss \citep{huang2021evidential} ($\mathcal{L}_{U}$), calibrated uncertainty loss \citep{zou2023evidencecap} ($\mathcal{L}_{CU}$) and a variant of calibrated uncertainty loss \citep{zou2023evidencecap} ($\mathcal{L}_{CU}(dis)$). Table \ref{tab:loss exp} gives the segmentation performance of models trained under different loss functions. On the AutoPET dataset, compared to the  model trained under the baseline loss $\mathcal{L}_{BS}$, models trained under $\mathcal{L}_{BS}+\mathcal{L}_{U}$ \citep{huang2021evidential}, $\mathcal{L}_{BS}+\mathcal{L}_{CU}$ \citep{zou2023evidencecap} or $\mathcal{L}_{BS}+\mathcal{L}_{CU}(dis)$ \citep{zou2023evidencecap} yield slight improvements in segmentation performance (77.92$\%$ vs. 78.65$\%$ vs. 77.93$\%$ vs. 78.46$\%$ in DSC). Specifically, when incorporating our proposed Uncertainty Perceptual Loss, the overall segmentation performance is significantly improved compared to that of the one trained under the baseline loss $\mathcal{L}_{BS}$ (77.92$\%$ vs. 82.45$\%$ in DSC). On the Hecktor dataset, although   the model trained under $\mathcal{L}_{BS}+\mathcal{L}_{UP}$ does not perform the best in terms of HD95 and precision, there is still a substantial improvement in overall metrics compared to the one trained under the baseline loss $\mathcal{L}_{BS}$. Fig. \ref{fig:loss exp} illustrates  changes in DSC with iterations during the training processes of models  under different loss functions. It can be observed from Fig. \ref{fig:loss exp} that the DSC of the model trained under $\mathcal{L}_{BS}$ shows a decreasing trend with increased iterations, indicating overfitting. After incorporating $\mathcal{L}_{UP}$ to the loss function, not only does the overall segmentation performance improve, but the  overfitting problem in the later training stages is also alleviated. With the incorporation of $\mathcal{L}_{U}$ \citep{huang2021evidential}, the model exhibits strong oscillations in the early training stages, which is possibly due to inaccurate quantification of uncertainty by the model in  initial stages. However, upon  introducing an annealing factor $\beta_t$, our proposed $\mathcal{L}_{UP}$ stabilizes the early training of the model.
\label{sec:Anylysis on Uncertainty Perceptual Loss}

\begin{figure}[t]
    \centering
    \includegraphics[width=0.6\columnwidth,trim=0 0 0 0,clip]{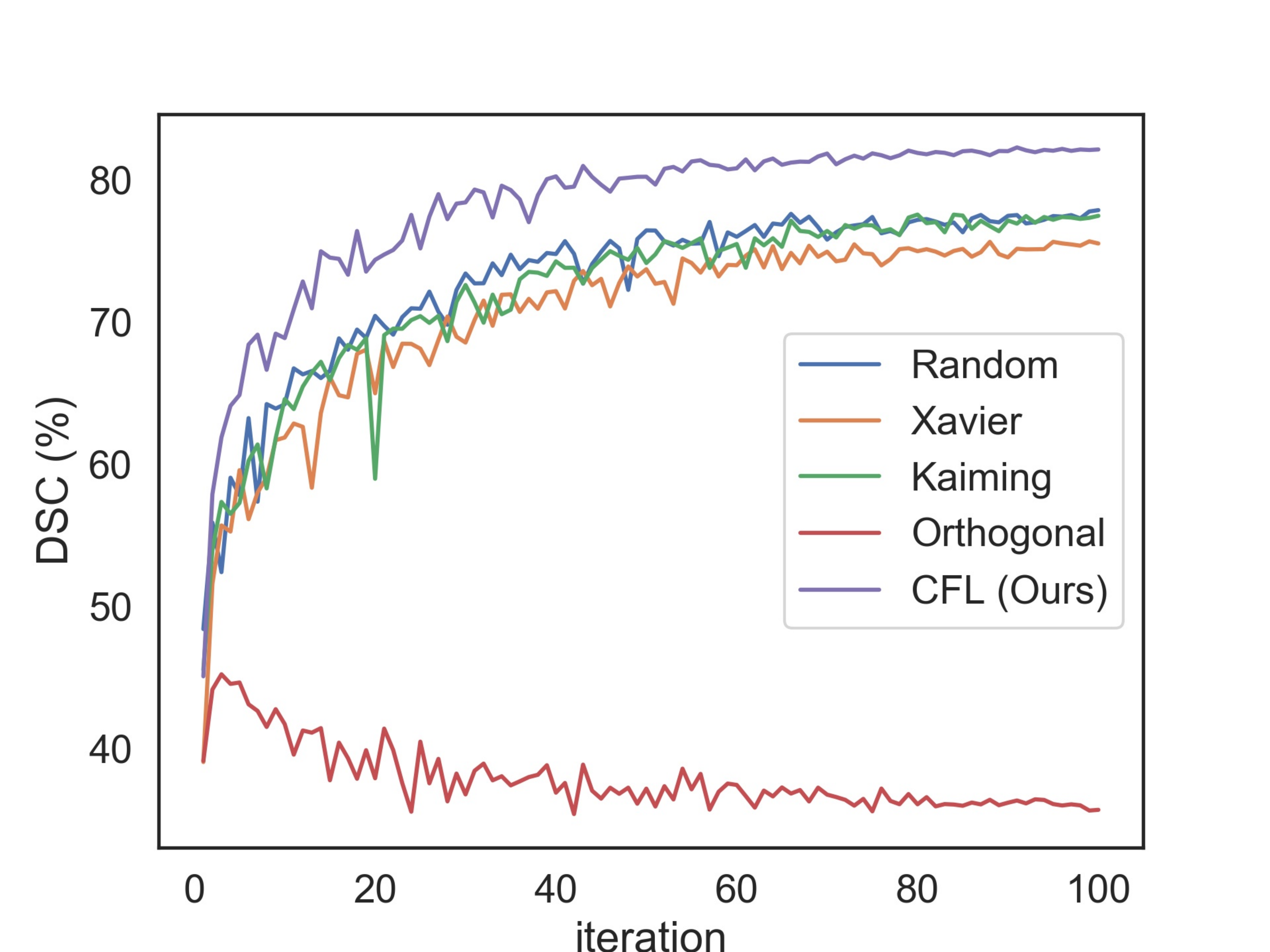}
    \caption{The performance comparison between the proposed CFL stage and four common initialization methods that do not use alternating training on the AutoPET dataset.}
    \label{fig:init exp}
\end{figure}

\subsubsection{Effectiveness of CFL}
In our work, a novel modality translation stage, called CFL, is introduced for alternate training and provides initialization parameters for the model in the MTF stage. To validate the effectiveness of CFL, we compare our method with four common initialization methods that do not use alternating training, and evaluate the model's training progress and performance using DSC.
These four initialization methods considered are Random (standard normal distribution), Xavier \citep{Xavier}, Kaiming \citep{he2015} and Orthogonal \citep{ahmad2023}. Fig. \ref{fig:init exp} illustrates curves of the DSC on the AutoPET dataset, while Table \ref{tab:init exp} provides the final DSC for both AutoPET and Hecktor datasets.

As a baseline,  the Random initialization strategy achieves final DSC of 77.87$\%$ and 82.28$\%$ on the AutoPET and Hecktor datasets, respectively. When using CFL, DSC on the same dataset are improved to 82.45$\%$ and 83.35$\%$, respectively, which demonstrates a significant improvement. Conversely, other initialization methods show minor improvements or even performance degradation. From the Fig. \ref{fig:init exp}, it can be clearly seen that when using Orthogonal \citep{saxe2013} initialization method, the model have fallen into a local optimum, resulting in poor generalization performance. It is evident that curves for the DSC when utilizing the CFL stage are notably higher than those with other initialization methods and exhibit smoother trends. These results convincingly demonstrate the effectiveness of CFL stage.
\label{sec:Analysis on initialize}
\label{sec:Ablation study}

\subsection{Visual assessment on segmentation uncertainty}
\begin{figure*}[t!]
    \centering
    \includegraphics[width=0.8\textwidth,trim=0 0 0 0,clip]{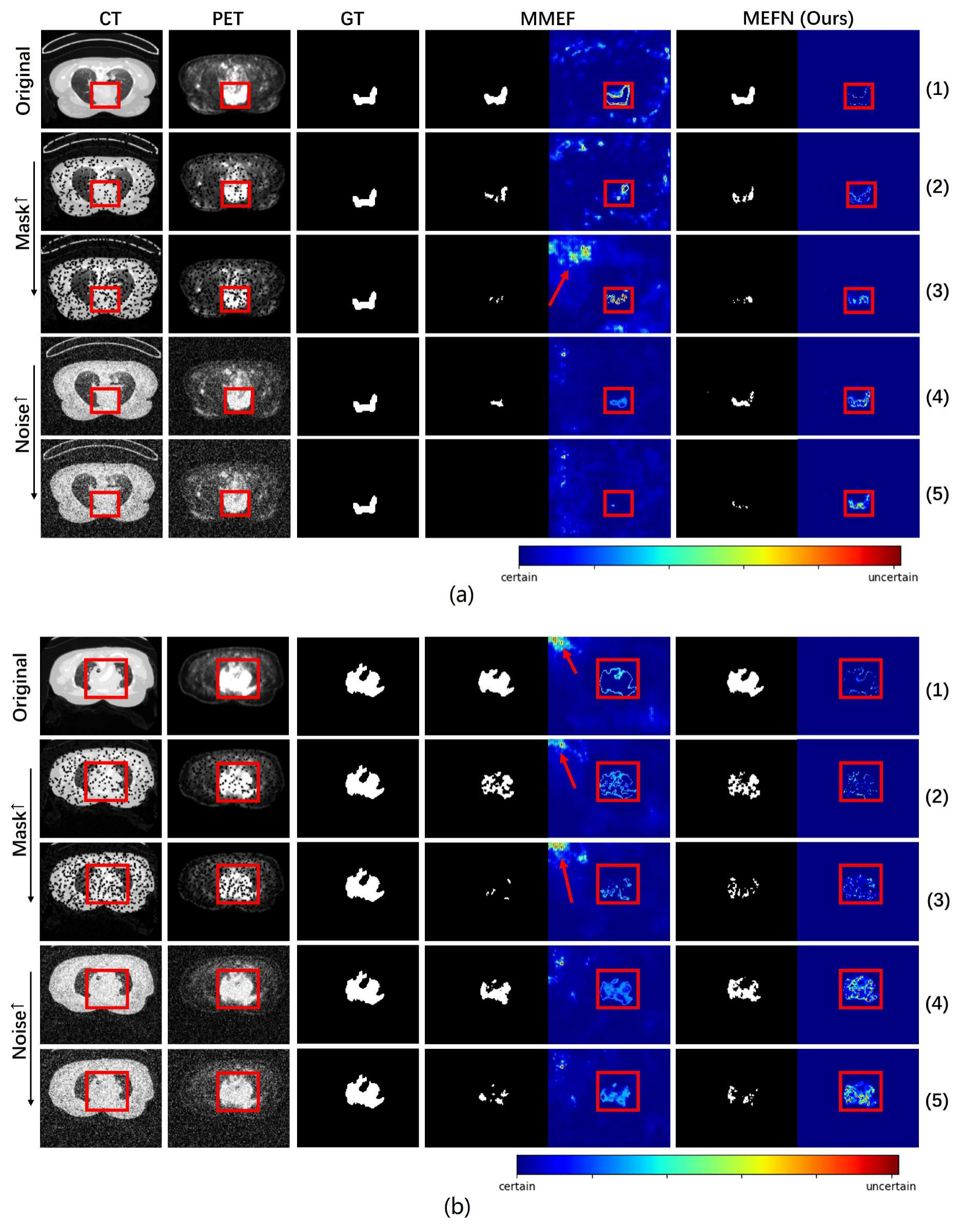}
    \caption{\justifying Segmentation results and their uncertainty for two sample images from the AutoPET dataset. Each row corresponds to the results for the  sample image or for their perturbed versions (with Gaussian noise ($\sigma^2 = 0.2, 0.3$) or patch-size random masked ($\gamma=0.04, 0.08$)). The columns correspond from left to right to sample CT image, sample PET image, ground truth labels, segmentation results and their uncertainty from MMEF and our MEFN, respectively.}
    \label{fig:uncertainty quantification exp}
\end{figure*}

 Intuitionally, the uncertainty output from different models should be positively correlated to data perturbation (i.e., the output uncertainty should increase with the incresement of perturbation levels of  input data). The conservation of such a positive correlation is a measure of the confidential level of the output uncertainty of different models. Thus,  we introduce perturbation of  different levels into the dataset with the same perturbation methods as described in Section \ref{sec:Analysis on Uncertainty Calibrator} and then conduct segmentation with different models. The correlation of the output uncertainty with the level of perturbation is employed to assess the reliability of the output uncertainty. We visually compare the  uncertainty output from the DTF module  of our MEFN  with that output from MMEF \citep{huang2021} for the segmentation results. Fig. \ref{fig:uncertainty quantification exp} gives the segmentation results of two sample images from the AutoPET dataset along with their segmentation uncertainty. The top part of this figure correspond to the first sample while the bottom part correspond to the second sample, with columns from  left to right corresponding to sample CT image, sample PET image, ground truth label, segmentation results from the MMEF and their uncertainty, segmentation results from our MEFN and their uncertainty. Rows of this figure correspond to the sample images themselves and and their perturbed versions.  Tumor regions are marked with red boxes.  It can be evidently seen from   the 4th and 6th columns of both samples that the segmentation results from both models become inferiors to different extent as the level of perturbation and the degree of Out-Of-Distribution (OOD) data increase.  Concurrently, the uncertainty predicted by our MEFN also consistently increases (as shown in the 7th columns in both samples), which implies a positive correlation between the uncertainty and the level of perturbation. In contrast, the uncertainty predicted with MMEF \citep{huang2021} inconsistently decreases (as shown in the 5th columns in both samples)  in some cases, which obviously violates the positive correlation requirement and is probably due to overconfidence in the model. These observations indicate that our proposed MEFN avoids blind confidence and provides radiologists and doctors with more credible uncertainty of the segmentation results. Moreover, beside tumor regions, MMEF \citep{huang2021} exhibits significant uncertainty in non-tumor regions (as pointed by red arrows in the uncertainty images in the third row in the first sample and first, seconde and third rows in the second sample). In contrast, MEFN only shows uncertainty in tumor regions(as shown in the 7th columns in both samples), indicating precise localization ability for tumors and more accurate handling of uncertainty in non-tumor regions. Furthermore, as shown in the 5th and 7th columns of the first rows  in both samples, our MEFN shows less uncertainty as compared with MMEF \citep{huang2021}, which indicates that our MEFN can provide more reliable segmentation results.

Besides, it can be observed from the last rows for both sample images that although the proposed method may not segment tumor regions well  for OOD data, it provides high uncertainty values in areas where tumors may exist, which prompts doctors performing further manual inspection of the tumorous regions. This can greatly boost doctors' confidence in trusting the model during clinical diagnosis, breaking the awkward situation of being hesitant to use deep learning due to its black-box nature.
\label{sec:Visualization and uncertainty quantification study}
\label{sec:Experiments and discussions}

\section{Conclusion}
This paper proposes a novel multi-modal Evidential fusion network for PET/CT tumor segmentation. The motivation of this work is to develop a method that can efficient and accurate fuse PET metabolic information and CT anatomical information. In this method, a GAN-based Cross-modality Feature Learning stage is first designed to align features across different modalities by mapping them to a shared feature space, thereby learning more robust feature representations. Then, two U-shaped networks are utilized to extract CT and PET features for feature fusion, during which a Dual-attention Calibrating module is specifically designed to reduce the impact of redundancy and confliction between features of different modalities on segmentation tasks. Finally, the output results are trustworthily fused based on the Dempster-Shafer Theory according to uncertainty. Extensive experiments are performed on two publicly available datasets whose results demonstrate the superiority of the proposed MEFN method outperforms other state-of-the-art multi-modal image segmentation methods. The improvements achieved with our MEFN method over other state-of-the-art methods in PET/CT multi-modal tumor segmentation amount to 3.10$\%$ and 3.23$\%$ in DSC on the AutoPET dataset and the Hecktor dataset, respectively.

A limitation of this work is that it is an end-to-end network model, resulting in large amount of model parameters and long training time. In future work, we will focus on deriving lightweight version of this network model or use two-stage training scheme to reduce network parameters while maintaining performance, making it more applicable in real clinical scenarios.
\label{sec:Conclusion}

\section*{CRediT authorship contribution statement}
\textbf{Yuxuan Qi:} Conceptualization, Writing - review and editing, Writing - original draft, Visualization, Software, Methodology, Investigation, Data curation, Formal Analysis. \textbf{Li Lin:} Conceptualization, Writing - review and editing, Methodology, Investigation, Supervision, Validation, Formal Analysis. \textbf{Jingya Zhang:} Resource, Data curation, Funding acquisition. \textbf{Jiajun Wang:} Conceptualization, Writing - review and editing, Validation, Supervision, Project administration, Funding acquisition. \textbf{Bin Zhang:} Resource, Investigation, Data curation, Funding acquisition.

\section*{Declaration of competing interest}
The authors declare that they have no known competing financial interests or personal relationships that could have appeared to influence the work reported in this paper.

\section*{Data availability}
Data will be made available on request.

\section*{Acknowledgments}
This work is supported by the National Natural Science Foundation of China, No. 60871086 and No. 61473243, the Natural Science Foundation of Jiangsu Province China, No. BK2008159 and the Natural Science Foundation of Suzhou No. SYG201113. The authors thank the anonymous reviewers for their constructive comments and valuable suggestions.

\bibliographystyle{elsarticle-num}
\bibliography{cas-refs}

\begin{thebibliography}{10}
\expandafter\ifx\csname url\endcsname\relax
  \def\url#1{\texttt{#1}}\fi
\expandafter\ifx\csname urlprefix\endcsname\relax\def\urlprefix{URL }\fi
\expandafter\ifx\csname href\endcsname\relax
  \def\href#1#2{#2} \def\path#1{#1}\fi

\bibitem{li2018}
X.~Li, H.~Chen, X.~Qi, Q.~Dou, C.-W. Fu, P.-A. Heng, H-denseunet: Hybrid densely connected unet for liver and tumor segmentation from ct volumes, IEEE Transactions on Medical Imaging 37~(12) (2018) 2663--2674.
\newblock \href {https://doi.org/10.1109/TMI.2018.2845918} {\path{doi:10.1109/TMI.2018.2845918}}.

\bibitem{yu2019}
Q.~Yu, Y.~Shi, J.~Sun, Y.~Gao, J.~Zhu, Y.~Dai, Crossbar-net: A novel convolutional neural network for kidney tumor segmentation in ct images, IEEE transactions on image processing 28~(8) (2019) 4060--4074.
\newblock \href {https://doi.org/10.1109/TIP.2019.2905537} {\path{doi:10.1109/TIP.2019.2905537}}.

\bibitem{ZHU2024}
Z.~Zhu, Z.~Wang, G.~Qi, N.~Mazur, P.~Yang, Y.~Liu, Brain tumor segmentation in mri with multi-modality spatial information enhancement and boundary shape correction, Pattern Recognition 153 (2024) 110553.
\newblock \href {https://doi.org/https://doi.org/10.1016/j.patcog.2024.110553} {\path{doi:https://doi.org/10.1016/j.patcog.2024.110553}}.

\bibitem{zhang2015}
W.~Zhang, R.~Li, H.~Deng, L.~Wang, W.~Lin, S.~Ji, D.~Shen, Deep convolutional neural networks for multi-modality isointense infant brain image segmentation, NeuroImage 108 (2015) 214--224.
\newblock \href {https://doi.org/https://doi.org/10.1016/j.neuroimage.2014.12.061} {\path{doi:https://doi.org/10.1016/j.neuroimage.2014.12.061}}.

\bibitem{chen2018}
H.~Chen, Q.~Dou, L.~Yu, J.~Qin, P.-A. Heng, Voxresnet: Deep voxelwise residual networks for brain segmentation from 3d mr images, NeuroImage 170 (2018) 446--455.
\newblock \href {https://doi.org/https://doi.org/10.1016/j.neuroimage.2017.04.041} {\path{doi:https://doi.org/10.1016/j.neuroimage.2017.04.041}}.

\bibitem{LI2023}
Y.~Li, S.~Li, H.~Ju, T.~Harada, H.~Zhang, T.~Duan, G.~Wang, L.~Zhang, L.~Gu, W.~Zhou, Correlated and individual feature learning with contrast-enhanced mr for malignancy characterization of hepatocellular carcinoma, Pattern Recognition 142 (2023) 109638.
\newblock \href {https://doi.org/https://doi.org/10.1016/j.patcog.2023.109638} {\path{doi:https://doi.org/10.1016/j.patcog.2023.109638}}.

\bibitem{nie2016}
D.~Nie, L.~Wang, Y.~Gao, D.~Shen, Fully convolutional networks for multi-modality isointense infant brain image segmentation, in: 2016 IEEE 13Th international symposium on biomedical imaging (ISBI), IEEE, 2016, pp. 1342--1345.

\bibitem{GUARRASI2024}
V.~Guarrasi, L.~Tronchin, D.~Albano, E.~Faiella, D.~Fazzini, D.~Santucci, P.~Soda, Multimodal explainability via latent shift applied to covid-19 stratification, Pattern Recognition 156 (2024) 110825.
\newblock \href {https://doi.org/https://doi.org/10.1016/j.patcog.2024.110825} {\path{doi:https://doi.org/10.1016/j.patcog.2024.110825}}.

\bibitem{goodfellow2020}
I.~Goodfellow, J.~Pouget-Abadie, M.~Mirza, B.~Xu, D.~Warde-Farley, S.~Ozair, A.~Courville, Y.~Bengio, Generative adversarial networks, Communications of the ACM 63~(11) (2020) 139--144.
\newblock \href {https://doi.org/https://doi.org/10.48550/arXiv.1406.2661} {\path{doi:https://doi.org/10.48550/arXiv.1406.2661}}.

\bibitem{wolterink2017}
J.~M. Wolterink, T.~Leiner, M.~A. Viergever, I.~I{\v{s}}gum, Generative adversarial networks for noise reduction in low-dose ct, IEEE transactions on medical imaging 36~(12) (2017) 2536--2545.
\newblock \href {https://doi.org/doi:10.1109/TMI.2017.2708987} {\path{doi:doi:10.1109/TMI.2017.2708987}}.

\bibitem{wang2018}
Y.~Wang, B.~Yu, L.~Wang, C.~Zu, D.~S. Lalush, W.~Lin, X.~Wu, J.~Zhou, D.~Shen, L.~Zhou, 3d conditional generative adversarial networks for high-quality pet image estimation at low dose, Neuroimage 174 (2018) 550--562.
\newblock \href {https://doi.org/https://doi.org/10.1016/j.neuroimage.2018.03.045} {\path{doi:https://doi.org/10.1016/j.neuroimage.2018.03.045}}.

\bibitem{bowles2018}
C.~Bowles, L.~Chen, R.~Guerrero, P.~Bentley, R.~Gunn, A.~Hammers, D.~A. Dickie, M.~V. Hern{\'a}ndez, J.~Wardlaw, D.~Rueckert, Gan augmentation: Augmenting training data using generative adversarial networks, arXiv preprint arXiv:1810.10863 (2018).
\newblock \href {https://doi.org/https://api.semanticscholar.org/CorpusID:53024682} {\path{doi:https://api.semanticscholar.org/CorpusID:53024682}}.

\bibitem{ben2019}
A.~Ben-Cohen, E.~Klang, S.~P. Raskin, S.~Soffer, S.~Ben-Haim, E.~Konen, M.~M. Amitai, H.~Greenspan, Cross-modality synthesis from ct to pet using fcn and gan networks for improved automated lesion detection, Engineering Applications of Artificial Intelligence 78 (2019) 186--194.
\newblock \href {https://doi.org/https://doi.org/10.1016/j.engappai.2018.11.013} {\path{doi:https://doi.org/10.1016/j.engappai.2018.11.013}}.

\bibitem{bi2017}
L.~Bi, J.~Kim, A.~Kumar, D.~Feng, M.~Fulham, Synthesis of positron emission tomography (pet) images via multi-channel generative adversarial networks (gans), in: Molecular Imaging, Reconstruction and Analysis of Moving Body Organs, and Stroke Imaging and Treatment: Fifth International Workshop, CMMI 2017, Second International Workshop, RAMBO 2017, and First International Workshop, SWITCH 2017, Held in Conjunction with MICCAI 2017, Qu{\'e}bec City, QC, Canada, September 14, 2017, Proceedings 5, Springer, 2017, pp. 43--51.

\bibitem{xiang2022}
X.~Zhang, B.~Zhang, S.~Deng, Q.~Meng, X.~Chen, D.~Xiang, Cross modality fusion for modality-specific lung tumor segmentation in pet-ct images, Physics in Medicine \& Biology 67~(22) (2022) 225006.
\newblock \href {https://doi.org/10.1088/1361-6560/ac994e} {\path{doi:10.1088/1361-6560/ac994e}}.

\bibitem{DING2024}
Y.~Ding, D.~Mu, J.~Zhang, Z.~Qin, L.~You, Z.~Qin, Y.~Guo, A cascaded framework with cross-modality transfer learning for whole heart segmentation, Pattern Recognition 147 (2024) 110088.
\newblock \href {https://doi.org/https://doi.org/10.1016/j.patcog.2023.110088} {\path{doi:https://doi.org/10.1016/j.patcog.2023.110088}}.

\bibitem{begoli2019}
E.~Begoli, T.~Bhattacharya, D.~Kusnezov, The need for uncertainty quantification in machine-assisted medical decision making, Nature Machine Intelligence 1~(1) (2019) 20--23.
\newblock \href {https://doi.org/10.1038/s42256-018-0004-1} {\path{doi:10.1038/s42256-018-0004-1}}.

\bibitem{gal2016}
Y.~Gal, Z.~Ghahramani, Dropout as a bayesian approximation: Representing model uncertainty in deep learning, in: international conference on machine learning, PMLR, 2016, pp. 1050--1059.
\newblock \href {https://doi.org/https://proceedings.mlr.press/v48/gal16.html} {\path{doi:https://proceedings.mlr.press/v48/gal16.html}}.

\bibitem{lakshminarayanan2017}
B.~Lakshminarayanan, A.~Pritzel, C.~Blundell, Simple and scalable predictive uncertainty estimation using deep ensembles, Advances in neural information processing systems 30 (2017).
\newblock \href {https://doi.org/https://proceedings.neurips.cc/paper_files/paper/2017/file/9ef2ed4b7fd2c810847ffa5fa85bce38-Paper.pdf} {\path{doi:https://proceedings.neurips.cc/paper_files/paper/2017/file/9ef2ed4b7fd2c810847ffa5fa85bce38-Paper.pdf}}.

\bibitem{Amersfoort2020}
J.~Van~Amersfoort, L.~Smith, Y.~W. Teh, Y.~Gal, Uncertainty estimation using a single deep deterministic neural network, in: International conference on machine learning, PMLR, 2020, pp. 9690--9700.
\newblock \href {https://doi.org/https://proceedings.mlr.press/v119/van-amersfoort20a.html} {\path{doi:https://proceedings.mlr.press/v119/van-amersfoort20a.html}}.

\bibitem{huang2022}
L.~Huang, S.~Ruan, P.~Decazes, T.~Den{\oe}ux, Lymphoma segmentation from 3d pet-ct images using a deep evidential network, International Journal of Approximate Reasoning 149 (2022) 39--60.
\newblock \href {https://doi.org/https://doi.org/10.1016/j.ijar.2022.06.007} {\path{doi:https://doi.org/10.1016/j.ijar.2022.06.007}}.

\bibitem{kohl2018}
S.~Kohl, B.~Romera-Paredes, C.~Meyer, J.~De~Fauw, J.~R. Ledsam, K.~Maier-Hein, S.~Eslami, D.~Jimenez~Rezende, O.~Ronneberger, A probabilistic u-net for segmentation of ambiguous images, Advances in neural information processing systems 31 (2018).
\newblock \href {https://doi.org/https://proceedings.neurips.cc/paper_files/paper/2018/file/473447ac58e1cd7e96172575f48dca3b-Paper.pdf} {\path{doi:https://proceedings.neurips.cc/paper_files/paper/2018/file/473447ac58e1cd7e96172575f48dca3b-Paper.pdf}}.

\bibitem{kamnitsas2018}
K.~Kamnitsas, W.~Bai, E.~Ferrante, S.~McDonagh, M.~Sinclair, N.~Pawlowski, M.~Rajchl, M.~Lee, B.~Kainz, D.~Rueckert, et~al., Ensembles of multiple models and architectures for robust brain tumour segmentation, in: Brainlesion: Glioma, Multiple Sclerosis, Stroke and Traumatic Brain Injuries: Third International Workshop, BrainLes 2017, Held in Conjunction with MICCAI 2017, Quebec City, QC, Canada, September 14, 2017, Revised Selected Papers 3, Springer, 2018, pp. 450--462.
\newblock \href {https://doi.org/https://doi.org/10.1007/978-3-319-75238-9_38} {\path{doi:https://doi.org/10.1007/978-3-319-75238-9_38}}.

\bibitem{mukhoti2021}
J.~Mukhoti, J.~van Amersfoort, P.~H. Torr, Y.~Gal, Deep deterministic uncertainty for semantic segmentation, arXiv preprint arXiv:2111.00079 (2021).
\newblock \href {https://doi.org/https://doi.org/10.48550/arXiv.2111.00079} {\path{doi:https://doi.org/10.48550/arXiv.2111.00079}}.

\bibitem{qayyum2021}
A.~Qayyum, A.~Benzinou, M.~Mazher, M.~Abdel-Nasser, D.~Puig, Automatic segmentation of head and neck (h\&n) primary tumors in pet and ct images using 3d-inception-resnet model, in: 3D Head and Neck Tumor Segmentation in PET/CT Challenge, Springer, 2021, pp. 58--67.
\newblock \href {https://doi.org/https://link.springer.com/chapter/10.1007/978-3-030-98253-9_4} {\path{doi:https://link.springer.com/chapter/10.1007/978-3-030-98253-9_4}}.

\bibitem{wang2021}
J.~Wang, Y.~Peng, Y.~Guo, D.~Li, J.~Sun, Ccut-net: pixel-wise global context channel attention ut-net for head and neck tumor segmentation, in: 3D Head and Neck Tumor Segmentation in PET/CT Challenge, Springer, 2021, pp. 38--49.
\newblock \href {https://doi.org/https://doi.org/10.1007/978-3-030-98253-9_2} {\path{doi:https://doi.org/10.1007/978-3-030-98253-9_2}}.

\bibitem{liu2021}
T.~Liu, Y.~Su, J.~Zhang, T.~Wei, Z.~Xiao, 3d u-net applied to simple attention module for head and neck tumor segmentation in pet and ct images, in: 3D Head and Neck Tumor Segmentation in PET/CT Challenge, Springer, 2021, pp. 99--108.
\newblock \href {https://doi.org/https://doi.org/10.1007/978-3-030-98253-9_9} {\path{doi:https://doi.org/10.1007/978-3-030-98253-9_9}}.

\bibitem{ZHOU2023}
T.~Zhou, Feature fusion and latent feature learning guided brain tumor segmentation and missing modality recovery network, Pattern Recognition 141 (2023) 109665.
\newblock \href {https://doi.org/https://doi.org/10.1016/j.patcog.2023.109665} {\path{doi:https://doi.org/10.1016/j.patcog.2023.109665}}.

\bibitem{WANG2023Cascaded}
C.~Wang, H.~Wang, Cascaded feature fusion with multi-level self-attention mechanism for object detection, Pattern Recognition 138 (2023) 109377.
\newblock \href {https://doi.org/https://doi.org/10.1016/j.patcog.2023.109377} {\path{doi:https://doi.org/10.1016/j.patcog.2023.109377}}.

\bibitem{ahmad2023}
I.~Ahmad, Y.~Xia, H.~Cui, Z.~U. Islam, Aatsn: Anatomy aware tumor segmentation network for pet-ct volumes and images using a lightweight fusion-attention mechanism, Computers in Biology and Medicine 157 (2023) 106748.
\newblock \href {https://doi.org/https://doi.org/10.1016/j.compbiomed.2023.106748} {\path{doi:https://doi.org/10.1016/j.compbiomed.2023.106748}}.

\bibitem{huang2021}
L.~Huang, T.~Den{\oe}ux, D.~Tonnelet, P.~Decazes, S.~Ruan, Deep pet/ct fusion with dempster-shafer theory for lymphoma segmentation, in: Machine Learning in Medical Imaging: 12th International Workshop, MLMI 2021, Held in Conjunction with MICCAI 2021, Strasbourg, France, September 27, 2021, Proceedings 12, Springer, 2021, pp. 30--39.
\newblock \href {https://doi.org/https://doi.org/10.1007/978-3-030-87589-3_4} {\path{doi:https://doi.org/10.1007/978-3-030-87589-3_4}}.

\bibitem{Zhu2017}
P.~Isola, J.-Y. Zhu, T.~Zhou, A.~A. Efros, Image-to-image translation with conditional adversarial networks, in: Proceedings of the IEEE conference on computer vision and pattern recognition, 2017, pp. 1125--1134.

\bibitem{vaswani}
A.~Vaswani, N.~Shazeer, N.~Parmar, J.~Uszkoreit, L.~Jones, A.~N. Gomez, {\L}.~Kaiser, I.~Polosukhin, Attention is all you need, Advances in neural information processing systems 30 (2017).
\newblock \href {https://doi.org/https://proceedings.neurips.cc/paper_files/paper/2017/file/3f5ee243547dee91fbd053c1c4a845aa-Paper.pdf} {\path{doi:https://proceedings.neurips.cc/paper_files/paper/2017/file/3f5ee243547dee91fbd053c1c4a845aa-Paper.pdf}}.

\bibitem{jsang2018}
A.~Jsang, Subjective Logic: A formalism for reasoning under uncertainty, Springer Publishing Company, Incorporated, 2018.
\newblock \href {https://doi.org/https://dl.acm.org/doi/abs/10.5555/3279217} {\path{doi:https://dl.acm.org/doi/abs/10.5555/3279217}}.

\bibitem{gatidis2022whole}
S.~Gatidis, T.~Hepp, M.~Fr{\"u}h, C.~La~Foug{\`e}re, K.~Nikolaou, C.~Pfannenberg, B.~Sch{\"o}lkopf, T.~K{\"u}stner, C.~Cyran, D.~Rubin, A whole-body fdg-pet/ct dataset with manually annotated tumor lesions, Scientific Data 9~(1) (2022) 601.
\newblock \href {https://doi.org/https://doi.org/10.1038/s41597-022-01718-3} {\path{doi:https://doi.org/10.1038/s41597-022-01718-3}}.

\bibitem{Andrearczyk2022head}
V.~Oreiller, V.~Andrearczyk, M.~Jreige, S.~Boughdad, H.~Elhalawani, J.~Castelli, M.~Valli{\`e}res, S.~Zhu, J.~Xie, Y.~Peng, et~al., Head and neck tumor segmentation in pet/ct: the hecktor challenge, Medical image analysis 77 (2022) 102336.
\newblock \href {https://doi.org/https://doi.org/10.1016/j.media.2021.102336} {\path{doi:https://doi.org/10.1016/j.media.2021.102336}}.

\bibitem{paszke2019pytorch}
A.~Paszke, S.~Gross, F.~Massa, A.~Lerer, J.~Bradbury, G.~Chanan, T.~Killeen, Z.~Lin, N.~Gimelshein, L.~Antiga, et~al., Pytorch: An imperative style, high-performance deep learning library, Advances in neural information processing systems 32 (2019).
\newblock \href {https://doi.org/https://proceedings.neurips.cc/paper_files/paper/2019/file/bdbca288fee7f92f2bfa9f7012727740-Paper.pdf} {\path{doi:https://proceedings.neurips.cc/paper_files/paper/2019/file/bdbca288fee7f92f2bfa9f7012727740-Paper.pdf}}.

\bibitem{zhang2021}
D.~Zhang, G.~Huang, Q.~Zhang, J.~Han, J.~Han, Y.~Yu, Cross-modality deep feature learning for brain tumor segmentation, Pattern Recognition 110 (2021) 107562.
\newblock \href {https://doi.org/https://doi.org/10.1016/j.patcog.2020.107562} {\path{doi:https://doi.org/10.1016/j.patcog.2020.107562}}.

\bibitem{wang2023SCL-Net}
M.~Wang, H.~Jiang, T.~Shi, Y.-d. Yao, Scl-net: Structured collaborative learning for pet/ct based tumor segmentation, IEEE Journal of Biomedical and Health Informatics 27~(2) (2023) 1048--1059.
\newblock \href {https://doi.org/10.1109/JBHI.2022.3226475} {\path{doi:10.1109/JBHI.2022.3226475}}.

\bibitem{wang2023}
F.~Wang, C.~Cheng, W.~Cao, Z.~Wu, H.~Wang, W.~Wei, Z.~Yan, Z.~Liu, Mfcnet: A multi-modal fusion and calibration networks for 3d pancreas tumor segmentation on pet-ct images, Computers in Biology and Medicine 155 (2023) 106657.
\newblock \href {https://doi.org/https://doi.org/10.1016/j.compbiomed.2023.106657} {\path{doi:https://doi.org/10.1016/j.compbiomed.2023.106657}}.

\bibitem{marinov2023}
Z.~Marinov, S.~Rei{\ss}, D.~Kersting, J.~Kleesiek, R.~Stiefelhagen, Mirror u-net: Marrying multimodal fission with multi-task learning for semantic segmentation in medical imaging, in: Proceedings of the IEEE/CVF International Conference on Computer Vision, 2023, pp. 2283--2293.

\bibitem{huang2021evidential}
L.~Huang, S.~Ruan, P.~Decazes, T.~Denoeux, Evidential segmentation of 3d pet/ct images, in: International conference on belief functions, Springer, 2021, pp. 159--167.
\newblock \href {https://doi.org/https://link.springer.com/chapter/10.1007/978-3-030-88601-1_16} {\path{doi:https://link.springer.com/chapter/10.1007/978-3-030-88601-1_16}}.

\bibitem{zou2023evidencecap}
K.~Zou, X.~Yuan, X.~Shen, Y.~Chen, M.~Wang, R.~S.~M. Goh, Y.~Liu, H.~Fu, Evidencecap: towards trustworthy medical image segmentation via evidential identity cap, arXiv preprint arXiv:2301.00349 (2023).

\bibitem{Xavier}
X.~Glorot, Y.~Bengio, Understanding the difficulty of training deep feedforward neural networks, in: Proceedings of the thirteenth international conference on artificial intelligence and statistics, JMLR Workshop and Conference Proceedings, 2010, pp. 249--256.
\newblock \href {https://doi.org/https://proceedings.mlr.press/v9/glorot10a.html} {\path{doi:https://proceedings.mlr.press/v9/glorot10a.html}}.

\bibitem{he2015}
K.~He, X.~Zhang, S.~Ren, J.~Sun, Delving deep into rectifiers: Surpassing human-level performance on imagenet classification, in: Proceedings of the IEEE international conference on computer vision, 2015, pp. 1026--1034.
\newblock \href {https://doi.org/https://openaccess.thecvf.com/content_iccv_2015/html/He_Delving_Deep_into_ICCV_2015_paper.html} {\path{doi:https://openaccess.thecvf.com/content_iccv_2015/html/He_Delving_Deep_into_ICCV_2015_paper.html}}.

\bibitem{saxe2013}
A.~M. Saxe, J.~L. McClelland, S.~Ganguli, Exact solutions to the nonlinear dynamics of learning in deep linear neural networks, arXiv preprint arXiv:1312.6120 (2013).
\newblock \href {https://doi.org/https://doi.org/10.48550/arXiv.1312.6120} {\path{doi:https://doi.org/10.48550/arXiv.1312.6120}}.

\end{thebibliography}

\end{document}